\documentclass[aps,12pt,nofootinbib,amsmath,eqsecnum, floatfix]{revtex4}
\usepackage{mathrsfs}
\usepackage{CJK}
\usepackage{amsmath}
\usepackage{amsfonts}
\usepackage{amssymb}
\usepackage{color}
\usepackage{undertilde}
\usepackage{ifpdf}

\ifpdf   
  \RequirePackage[pdftex]{hyperref}
\else    
    \RequirePackage[dvips]{hyperref}
  \fi
\hypersetup{bookmarksnumbered,%
              colorlinks,%
               linkcolor=blue,%
               citecolor=blue,%
              plainpages=false,%
            pdfstartview=FitH}

\def\no{\noindent}

\def\bc{\begin{center}}
\def\nno{\nonumber}
\def\ec{\end{center}}
\def\be{\begin{eqnarray}}
\def\ee{\end{eqnarray}}

\newcommand{\omits}[1]{}

\definecolor{dyellow}{rgb}{1.,0.8,.0}
\definecolor{myblue}{rgb}{.1,.1,.7}
\definecolor{dcyan}{rgb}{.0,.6,.6}
\definecolor{dmagenta}{rgb}{0.6,0.0,0.6}
\definecolor{brown}{rgb}{0.6,0.2,0.}
\definecolor{darkblue}{rgb}{.0,.0,0.5}
\definecolor{darkred}{rgb}{0.75,0.0,0.0}
\definecolor{orange}{rgb}{1.,.6,.0}
\definecolor{dorange}{rgb}{0.8,.4,.0}
\definecolor{darkgreen}{rgb}{0.0,0.6,0.0}
\definecolor{purple}{rgb}{.4,.0,.4}
\definecolor{lightgrey}{rgb}{0.7,0.7,0.7}

\def\black{\color{black}}

\def\blue{\color{blue}}

\def\Ga{\Gamma}

\def\Si{\Sigma}
\def\Om{\Omega}
\def\al{\alpha}
\def\ga{\gamma}
\def\dl{\delta}
\def\eps{\epsilon}
\def\ka{\kappa}
\def\la{\lambda}
\def\th{\theta}
\def\si{\sigma}

\def\del{\nabla}
\def\d#1#2{\frac{\displaystyle #1}{\displaystyle #2}}
\def\r{\partial}

\newcommand{\dS}{$d{S}$}

\newcommand{\Mink}{Minkowski}

\newcommand{\PoR}{${P}o{ R}$}
\newcommand{\PoRcl}{${P}o{R}_{c,l}$}

\newcommand{\vect}[1]{\mbox{\boldmath $#1$}}

\newcommand\btd{\raise 2pt
\hbox{$\hat\bigtriangledown$}\hskip 1.5pt}
\newcommand\bt{\raise 2pt
\hbox{$\bigtriangledown$}\hskip 1.5pt}

\newtheorem{thm}{Theorem}

\begin{document}
\begin{CJK*}{GBK}{song}


\title{Geometries with the second Poincar\'e symmetry}%

\author{{Chao-Guang Huang}$^{1}$} \email{huangcg@ihep.ac.cn}
\author{{Yu Tian}$^2$}\email{ytian@gucas.ac.cn}
\author{{Xiao-Ning Wu}$^{3}$}\email{wuxn@amss.ac.cn}
\author{{Zhan Xu}$^{4}$}\email{zx-dmp@tsinghua.edu.cn}
\author{{Bin Zhou}$^{5}$} \email{zhoub@bnu.edu.cn}

\affiliation{%
${}^1$ Institute of High Energy Physics, and
Theoretical Physics Center for Science Facilities, Chinese Academy of
Sciences, Beijing 100049, China}

\affiliation{%
${}^2$ Graduate University of Chinese Academy of
Sciences, Beijing 100049, China}

\affiliation{%
${}^3$Institute of Mathematics, Academy of Mathematics and System Sciences,
Chinese Academy of Sciences, Beijing
100190, China,}

\affiliation{
${}^4$ Department of Physics, Tsinghua University, Beijing
100084, China}

\affiliation{%
${}^5$ Department of Physics, Beijing Normal University, Beijing
100875, China.}

\begin{abstract}
The second Poincar\'e kinematical group serves as one of new
ones in addition to the known possible kinematics.  The
geometries with the second Poincar\'e symmetry is presented and
their properties are analyzed.  On the
geometries, the new mechanics based on the
principle of relativity with two universal constants
$(c,l)$ can be established.

\bigskip

\no

\no

\end{abstract}

\maketitle

\tableofcontents


\section{Introduction}

\black

It is well known that a maximum symmetry group of a 4d non-degenerate
space-time has 10 independent parameters.  Minkowski (Mink), de
Sitter ($dS$), and anti de Sitter ($AdS$) space-times are all the
space-times of this kind.  In addition, the Galilei ($G$) space-time
has 10-parameter kinematical group even though its geometry is
degenerate, splitting into 3d space and 1d time geometries. It is
natural to ask: how many are there 10-parameter kinematical groups
in 4d space-time? Bacry and L\'evy-Leblond have
answered the question
\cite{BLL}. Under their three assumptions and by the
In\"on\"u-Wigner contraction method \cite{IW}, they show that there
are 8 types of Lie algebras corresponding to 11 possible kinematical
groups. They are the Poincar\'e ($P$), $dS$, $AdS$, inhomogenous
SO(4) ($P'_+$), para-Poincar\'e ($P'_-$), $G$, Newton-Hooke
($NH_+$), anti-Newton-Hooke ($NH_-$), para-Galilei ($G'$), Carroll
($C$), and static ($S$) groups.  If their third assumption that inertial
transformations in any given direction form a noncompact subgroup
is relaxed, other 3 classical, geometrically kinematical groups will be added,
corresponding to Euclid ($Euc$), Riemann ($Riem$), and Lobachevski
($Lob$) geometries.

On the other hand, the principle of relativity (\PoR) is the foundation of
physics, and it is closely related to the symmetry of space and
time.  Recently, it is pointed out that the triality of special
relativity with Poincar\'e, de Sitter, anti-de Sitter invariance, respectively,
can be set up based on the \PoR\ and the postulate on two universal
invariant constants $c$ of speed dimension and $l$ of length
dimension denoted as the $PoR_{c,l}$ \cite{GWZ, GHWZ}.  It is also
found in \cite{GWZ, GHWZ} that there is another realization of Poincar\'e group being called
the second
Poincar\'e group and denoted as $P_2$, with the corresponding
realization of algebra being denoted as
${\frak p}_2$.\footnote{For brevity, we call the new realization of the Poincar\'e
group as well as its algebra the second Poincar\'e group and the second Poincar\'e algebra
throughout the paper.} \omits{, distinguished from the ordinary
realization of Poincar\'e group.}\omits{{\blue The algebra can be called
the (second) Poincar\'e algebra} in the following algebraic sense.  The
unique Abelean ideal of the $\frak{iso}(1,3)$ algebra serves as
translation sub-algebra and is divided into the time translation and
space translations
as a 1d and a 3d representation, respectively, of
$\mathfrak{so}(3)$ sub-algebra of the $\frak{so}(1,3)$ sub-algebra, no matter
whether the translation generators generate genuine translations in space-time.
In addition, the algebra is invariant under the suitably defined parity and time-reversal
operation \cite{BLL}.}  Unlike the ordinary Poincar\'e
transformation under which the metric of Minkowski space-time is
invariant, the second Poincar\'e transformations do not generate the
automorphism of the Minkowski space-time. Instead,
they preserve all straight lines in the Minkowski
space-time.  Furthermore, it has been shown based on the
$PoR_{c,l}$ that every algebra in all possible kinematics revealed
by Bacry and L\'evy-Leblond except 4 semi-simple groups, $dS$,
$AdS$, $Riem$, and $Lob$ groups has its second version
\cite{GHWZ}.  Therefore, there are 24 kinematical groups with
$SO(3)$ isotropic subgroup in all.  One of the reasons of
the absence of the second versions of many groups, such as the second
Poincar\'e group, in \cite{BLL} is that they just consider the
algebraic structure but not consider the action space of the group. A
natural question appears: what are the meanings of these additional
possible kinematical algebras or what do these additional possible
kinematical algebras represent?

In order to clarify the question, one has to know on what kinds of
4d space-times these possible kinematics are defined.
Unfortunately, more than a half of the space-times corresponding to
these kinematical algebras are unknown.  Our recent letter ameliorated the situation
somewhat, in which we presented a geometry with the
$P_2$ symmetry\cite{lett}.  One of the purposes of the paper is to make the thorough
investigation on the $P_2$ geometries.

Both in the treatment of Bacry and L\'evy-Leblond and the approach
based on $PoR_{c,l}$, which are very different from each other,  $SO(3)$
symmetry is identified as the  space isotropy in algebraic sense.
However, the sub-algebras in the
possible kinematical algebras can be interpreted in many ways.
Whether an $SO(3)$ isotropy can be identified to the space
isotropy is determined by the geometrical (as well as topological and causal)
structure of the space-time.  In other
words, before a careful geometrical study, we cannot conclude that
the space-times possess the space $SO(3)$ isotropy in the
geometries even though the corresponding kinematical algebras
having $SO(3)$ sub-algebra. The second purpose of the paper is to
take the geometrical structure with $P_2$ symmetry as an example
to clarify that the algebraic $\frak{so}(3)$ isotropy in \cite{BLL, GHWZ} does not
always imply the geometrical $SO(3)$ space isotropy.

Once the topology and geometry are clarified, one needs to
re-construct the algebras according to the understanding of the
geometry.
The third purpose of the paper is to show that there exists the geometrical
structure which satisfies all three assumptions in
\cite{BLL} and the \PoRcl\ in \cite{GWZ, GHWZ} even after re-construction of the
algebra in terms of new space and time coordinates.

The paper will be organized in the following way.  In the next section, we shall
review the second Poincar\'e symmetries.  Section III focuses on the no-go theorem
that there does not exist a non-degenerate
metric with the second Poincar\'e symmetry.  In sections IV and V, we shall
present degenerate metrics and connection which are $P_2$ invariant and study
the structure of the space-times
described by the metrics and connection, respectively.  In section VI,
we prove the uniqueness of the geometrical structures for the second Poincar\'e
symmetry.  Then, we show that the maximum symmetry of the new space-times is the
Poincar\'e symmetry and re-classify the generators according to geometries in
section VII.
Section VIII is devoted to
set up the mechanics of a free particle on the geometry.
We shall study the uniform rectilinear motions in the space with degenerate
metrics and present the formal Lagrangian formalism for the particle moving
on the geometries.
Finally, we shall conclude the paper with some concluding remarks in section IX.


\section{The second Poincar\'e symmetry}

The ordinary Poincar\'e transformations%
\be \label{ptr}
{x'}^\mu = L^\mu_{\ \nu}x^\nu+ l a^\mu, \qquad L\in SO(1,3),
\ee
\omits{\be {x'}^\mu =(x^\nu-a^\nu)
L^\mu_{\ \nu}, \qquad L\in SO(1,3), \ee}%
where $a^\mu$ are dimensionless parameters, transfer the origin $O(o^\mu)$ to
the event $P(x_P^\mu= l (L^{-1})^\mu_{\ \nu} a^\nu)$
and a
generator set $\{T\}^{{\frak p}} =(H, \vect{P}, \vect{K},\vect{J})$ \footnote{$\vect{P}$,
$\vect{K}$ $\cdots$ are the shorthands of $P_i$, and $K_i$,  $\cdots$, respectively, where
$i=1,2,3$. }
spans a Poincar\'e algebra ${\frak p} \cong {\frak{iso}}(1,3)$,
\be\label{p}
H = \r_t ,\ P_i = \r_i,\ K_i = t \r_i +\d 1 {c^2} x^i \r_t,\
J_i = \eps_{i}^{\ jk}(x_j\r_k-x_k\r_j),
\ee%
where the indexes are lowered or raised by $(\eta_{\mu\nu})={\rm diag}(1,-1,-1,-1)$
and its inverse.  The transformation (\ref{ptr}) can be expressed in a $5\times 5$ matrix,
\be \label{pmat}
\left(
  \begin{array}{cc}
    L & a \\
    0 & 1 \\
  \end{array}
\right)
\ee%
\black
With the same $\vect{K}, \vect{J}$, there exists a second generator set
$\{T\}^{{\frak p}_2} =(H',\vect{P} ', \vect{K},\vect{J})$, where
\be\label{np2}
H'  = - c^2 l^{-2} t  x^\ka \r_\ka  \ (=c P_0'), \qquad
 P'_i  = l^{-2} x^i  x^\ka \r_\ka .
\ee
They spans the second Poincar\'e algebra ${\frak p}_2$,
\begin{equation}
\begin{split}
&[ H', P'_i ]=0,  \qquad  [ P'_i, P'_j] = 0, \qquad
[H', K_i]= P'_i, \qquad  [ P'_i, K_j]=
\d 1 {c^2} H'\delta_{ij},
\\
&[ K_i, K_j ] = - \d 1 {c^2}L_{ij} \quad [J_i, J_j]=
-\eps_{ij}^{\ \ k}J_k .
\end{split}
\end{equation}
In other words, there is no difference between the ordinary Poincar\'e
algebra and the second Poincar\'e algebra in {\it algebraic} sense.
However, the second Poincar\'e algebra is the different realization of $\frak{iso}(1,3)$
from the ordinary realization.  The second Poincar\'e algebra
generates the second Poincar\'e transformations
\be\label{CoordTrans}
{x'}^\mu = \d {L^\mu_{\ \nu}x^\nu}{1+l^{-1}b_\la x^\la} ,
\ee
where $b_\mu$ are dimensionless parameters, which can be expressed again in terms
of $5\times 5$ matrix
\be \label{p2mat}
\left(
  \begin{array}{cc}
    L & 0 \\
    b^t & 1 \\
  \end{array}
\right)
\ee%
where $b^t:=(\eta_{\mu\nu} b^\nu)$ is the transpose of
$4\times 1$ matrix $b$. Clearly, as a part of linear fractional
transformations, they preserve all straight lines,
\be
\begin{cases}
x^0=ct, \\
x^i=v^i t + x_0^i,
\end{cases} \qquad \mbox{$v^i$, $x_0^i$ are arbitrary constants, }
\ee
no matter whether the lines are causal ($c^2-\dl_{ij}v^i v^j \geq 0$) or not.
In particular, they preserve the light cone at the origin
\be %
\eta_{\mu\nu} x^\mu x^\nu=0.
\ee
A simple calculation shows that the second Poincar\'e
transformations do not preserve the metric  of the \Mink\
space-time.\footnote{It should be
noted that  the second Poincar\'e group presented here is different
from the second Poincar\'e group presented by Aldrovandi and Pereira \cite{AP}.
The second Poincar\'e group here is the semi-direct product of the pseudo-translations and
Lorentz group and is a subgroup of the general projective group,
while the second Poincar\'e group presented by Aldrovandi and Pereira
is the semi-direct product of the special conformal transformations and Lorentz group
and is a subgroup of the conformal transformation
 group.}

To be distinguished from the ordinary time and
space translation generators $H$ and $\vect{P}$, $H'$ and $\vect{P}'$ are
called the pseudo-time- and pseudo-space-translation generators
because they cannot generate time or space translation in \Mink\
space-time.


\section{No-go theorem}

\begin{thm}
There is no tensor field $\vect{g} = g_{\mu\nu} \, dx^\mu \otimes dx^\nu$
with the following three conditions satisfied simultaneously:
(1) $\vect{g}$ is smooth;
(2) $\vect{g}$ is non-degenerate everywhere;
(3) $\vect{g}$ is invariant under the $\mathfrak{p}_2$-translations.
\end{thm}

If the theorem was incorrect, there would be certain a tensor field $\vect{g}$,
which would be treated as the metric, satisfying the conditions in the theorem.
Let $\nabla$ be the Levi-Civita connection related to $\vect{g}$. The Killing
equation for vector field ${P^\mu}^a$ denoted by
$- l^{-2}  x^\mu D^a$ with $D^a =D^\ka (\partial_\ka)^a= x^\ka (\partial_\ka)^a$
would read
\begin{equation}
 0 =\del_a(x^\mu D_b)+\del_b(x^\mu D_a) =
 x^\mu \, (\nabla_a D_b + \nabla_b D_a) + (dx^\mu)_a  D_b + D_a (dx^\mu)_b,
  \label{eq:Killing:P-mu}
\end{equation}
where $D_a = g_{ab} D^b=D_\mu (dx^\mu)_a$.  ($D_\mu =g_{\mu\nu}D^\nu$.)
The contraction of $D^b$ and $D_\mu$
with Eq. (\ref{eq:Killing:P-mu}), respectively, yield
\begin{eqnarray}
  x^\mu \, \Big(D^b \nabla_b D_a + \frac{1}{2} \nabla_a (D^b D_b) + D_a\Big)
  + (dx^\mu)_a  D^b D_b
  & = & 0
  \, , \label{eq:tmp:1} \\
  D_\mu D^\mu  \, (\nabla_a D_b + \nabla_b D_a) + 2 \, D_a D_b & = & 0
  \, .
  \label{eq:tmp:2}
\end{eqnarray}\black
The contraction of Eq.~(\ref{eq:tmp:1}) with $D_\mu$ gives rise to
\[
 D_\mu D^\mu \Big(
      D^b \nabla_b D_a + \frac{1}{2} \nabla_a (D^b D_b) + 2\, D_a
    \Big)
  = 0
  \, ,
\]
which is valid for an arbitrary point $p$, either $D_\mu D^\mu |_p = 0$ or
\begin{equation}
  \Big(D^b \nabla_b D_a + \frac{1}{2} \nabla_a (D^b D_b) + 2\, D_a\Big)
    \Big|_p
  = 0 \, .
  \label{eq:tmp:3}
\end{equation}
If $D_\mu D^\mu |_p = 0$, Eq.~(\ref{eq:tmp:2}) requires $D_a D_b
|_p = 0$, which implies that $D_a |_p = 0$. Since $g_{ab}$ is
non-degenerate, it is possible if and only if $D^a |_p = 0$. In
other words, it is possible if and only if $p$ is the origin of
the coordinate system. When $p$ is not the origin,
Eq.~(\ref{eq:tmp:3}) is always satisfied, together with $D^b D_b
|_p \neq 0$.  Then Eq.~(\ref{eq:tmp:1}) results in
\begin{equation}
  (dx^\mu)_a |_p = \frac{x^\mu}{D^d D_d} \, D_a \Big|_p
  \, ,
\end{equation}
which is absurd because $dx^0 |_p$, $dx^1 |_p$, $dx^2 |_p$ and $dx^3 |_p$
are linearly independent.  \hfill{$\Box$}


\section{Geometries for $P_2$ symmetry}

The no-go theorem shows that the $\mathfrak{p}_2$-invariant metrics on the 4d
underlying manifold must be degenerate.
In order to completely fix the geometry of the space-time with a degenerate metric,
more information should be assigned.

Consider a 4d manifold $M^{{\frak p}_2}$ endowed with (1) a type-(0,2) degenerate
symmetric tensor field\footnote{The abstract and component forms of tensor fields
are both used in the present paper.}
\be \label{g1}%
\vect{g}^\pm=g^\pm_{\mu\nu}dx^\mu \otimes dx^\nu= \pm \frac {l^2} {(x \cdot x)^2}
(\eta_{\mu \rho}\eta_{\nu \tau}-\eta_{\mu\nu}\eta_{\rho \tau})x^\rho x^\tau
dx^\mu dx^\nu,
\ee
where
\be
x \cdot x=
\eta_{\mu\nu} x^\mu x^\nu \begin{cases} < 0, & \mbox{for upper sign}\\
                                                  > 0, & \mbox{for lower sign}
                                                  \end{cases}%
\ee%
(2)
a type-(2,0) degenerate symmetric tensor field%
\be \label{gp2inv}
\vect{h}_\pm=h_\pm^{\mu\nu}\r_\mu \otimes \r_\nu =
l^{-4}(x\cdot x)x^\mu x^\nu\r_\mu \otimes \r_\nu
\ee%
and (3) a connection $\del^\pm$ compatible to ${\vect g}^\pm$ and ${\vect h}_\pm$,
i.e.%
\be
\del^\pm_\la \, g^\pm_{\mu\nu}= \r_\la
g^\pm_{\mu\nu}-\Gamma^\ka_{\la\nu}g^\pm_{\mu\ka}
-\Gamma^\ka_{\mu\la}g^\pm_{\ka\nu} =0
\ee%
and %
\be
\del^\pm_\la \,
{h}_\pm^{\mu\nu}= \r_\la {h}_\pm^{\mu\nu}+
\Gamma^\nu_{\la\ka}{h}_\pm^{\mu\ka}
+\Gamma^\mu_{\la\ka}{h}_\pm^{\ka\nu}=0,
\ee%
respectively, with connection coefficients in the above coordinate system,%
\be
\label{inv-connectn}
\Gamma^\mu_{\pm\, \nu\la} = - \d {x_\nu \dl^\mu_\la+\dl^\mu_\nu x_\la }{x\cdot x}.
\ee%
It is easy to
check that $(M^{{\frak p}_2}_\pm,\vect{g}^\pm,\vect{h}_\pm,\del^\pm)$ is
invariant under $P_2$ transformation, namely, $\forall \vect\xi
\in {\frak p}_2\subset \Gamma(TM^{\frak p_2})$, equations%
\be \label{eq:Lieg}
{\cal
L}_{\vect \xi} \vect{g}^\pm = (\xi^\la\r_\la g^\pm_{\mu\nu}+g^\pm_{\mu\la}\r_\nu\xi^\la
+g^\pm_{\la\nu}\r_\mu\xi^\la)dx^\mu\otimes dx^\nu =0,
\ee
\be \label{eq:Lieh}
{\cal L}_{\vect \xi} {\vect h}_\pm=(\xi^\la\r_\la{h}^{\mu\nu}_{\pm } -
h_\pm^{\mu\la}\r_\la\xi^\nu - h_\pm^{\la\nu}\r_\la\xi^\mu)\r_\mu\otimes\r_\nu=0,
\ee
and
\be \label{eq:LieGamma}
[{\cal L}_{\vect \xi}, \del^\pm ] =0
\ee
are valid simultaneously. In other words, Eqs.(\ref{g1}),
(\ref{gp2inv}), (\ref{inv-connectn}) are invariant under the
coordinate transformation (\ref{CoordTrans}) and its inverse
transformation,
\be \label{InverseTransf}%
x=\d {L^{-1} x'}{1-l^{-1}(b\cdot L^{-1}x')}=\d
{L^{-1} x'}{1-l^{-1}(b'\cdot x')}.
\ee

By definition, the curvature tensor is
\be
R^\si_{\pm\, \mu \nu \rho}=\r_\nu \Gamma^\si_{\pm\, \mu \rho  }-
\r_\rho\Gamma^\si_{\pm\, \mu \nu}
+\Gamma^{\si}_{\pm\, \tau \nu} \Gamma^\tau_{\pm\, \mu \rho}-
\Gamma^{\si}_{\pm\, \tau \rho} \Gamma^\tau_{\pm\, \mu \nu}=
\pm l^{-2}(\dl^\si_\rho g^\pm_{\mu\nu}
-\dl^\si_\nu g^\pm_{\mu\rho}). \label{curvaturecomponents}
\ee
It is antisymmetric in the latter two indexes and satisfies the Ricci and
Bianchi identities. The Ricci curvature tensor is then
\be \label{riccicurv}
R^\pm_{\mu\nu}=R^\si_{\pm\, \mu\nu\si}=\pm 3l^{-2}g^\pm_{\mu\nu}.
\ee
They are obviously invariant under $P_2$ transformation.  Eqs.(\ref{curvaturecomponents})
and (\ref{riccicurv}) are similar to those of the maximum-symmetric space-times.

\bigskip


\section{Structure of the space-times}

The geometries in the previous section are presented
in a special coordinate system $x^\mu$.
In order to see the structures of the manifolds more transparently, we consider the
coordinate transformations,
\be
\label{Transf2dS}
\begin{cases}
x^0=l^2\rho^{-1} \sinh (\psi/l) &\\
x^1=l^2\rho^{-1} \cosh (\psi/l) \sin \th \cos\phi&\\
x^2=l^2\rho^{-1} \cosh (\psi/l) \sin \th \sin \phi&\\
x^3=l^2\rho^{-1} \cosh (\psi/l) \cos \th
\end{cases}\qquad \mbox{for }(x\cdot x)<0,
\ee
\be \label{Transf2AdS}
\begin{cases}
x^0=l^2\eta^{-1} \cosh (r/l) &\\
x^1=l^2\eta^{-1} \sinh (r/l) \sin \th \cos\phi&\\
x^2=l^2\eta^{-1} \sinh (r/l) \sin \th \sin \phi&\\
x^3=l^2\eta^{-1} \sinh (r/l) \cos \th
\end{cases} \qquad \mbox{for }(x\cdot x)>0,
\ee
respectively. Under the coordinate transformations, Eqs.(\ref{g1}), (\ref{gp2inv}), and
(\ref{inv-connectn}) become, respectively,
\be \label{g3}
{\vect g}^\pm =
\bar g^\pm_{\mu\nu}d\bar x^\mu d\bar x^\nu=\begin{cases}d\psi^2 -l^2\cosh^2(\psi/l) d\Om_2^2
& {\rm for}\ x\cdot x <0\\
-d r^2 -l^2\sinh^2(r/l) d\Om_2^2& {\rm for}\ x\cdot
x>0,\end{cases}
\ee
\be\label{h1}
{\vect h}_\pm=\bar h_\pm^{\mu\nu}\r_\mu \otimes\r_\nu=\begin{cases}-
\d {\r }{\r \rho}\otimes \d {\r }{\r \rho}, & {\rm for}\ x\cdot x <0
\smallskip \\
 \d {\r }{\r \eta} \otimes \d {\r }{\r \eta}, &{\rm for}\ x\cdot x >0,
\end{cases}
\ee
\be \label{connect+}\begin{cases}
{\bar \Gamma}^\psi_{+\th\th}=l\sinh(\psi/l) \cosh(\psi/l),\quad
{\bar \Gamma}^\psi_{+\phi\phi}=
{\bar \Gamma}^\psi_{+\th\th}\sin^2\th &\\
{\bar \Gamma}^\th_{+\th\psi}={\bar \Gamma}^\th_{+\psi\th}={\bar \Gamma}^\phi_{+\phi\psi}
={\bar \Gamma}^\phi_{+\psi\phi}
=l^{-1} \tanh(\psi/l) &\\
{\bar \Gamma}^\th_{+\phi\phi}=- \sin \th \cos \th , \quad
{\bar \Gamma}^\phi_{+\th\phi}={\bar \Gamma}^\phi_{+\phi\th}= \cot \th &\\
{\bar \Gamma}^\rho_{+\al\beta} = -l^{-2}\rho g_{\al\beta}, \qquad
\mbox{others vanish},&\end{cases}\qquad {\rm for} \ x\cdot x <0,
\ee

\be\label{connect-}\begin{cases}
{\bar \Gamma}^\eta_{-ij} = +\ l^{-2}\eta g_{ij}&\\
{\bar \Gamma}^r_{-\th\th} =-l\sinh(r/l) \cosh(r/l), \quad
{\bar \Gamma}^r_{-\phi\phi} ={\bar \Gamma}^r_{-\th\th}\sin^2\th   &\\
{\bar \Gamma}^\th_{-r\th} ={\bar \Gamma}^\th_{-\th r} ={\bar \Gamma}^\phi_{-r\phi}
={\bar \Gamma}^\phi_{-\phi r}
=\d 1 {l\tanh(r/l)}& \\
{\bar \Gamma}^\th_{-\phi\phi} =-\sin\th\cos\th, \quad
{\bar \Gamma}^\phi_{-\th\phi}={\bar \Gamma}^\phi_{-\phi\th}=\cot\th& \\
\mbox{others vanish} &
\end{cases}  \qquad {\rm for}\ x\cdot x >0,%
\ee
where
\be
(\bar x^\mu)=\begin{cases}(\bar x^\al; \bar x^3)
=(\psi, \th, \phi; \rho), \quad \al,\ \beta, \ \ga \mbox{ run from 0 to 2,}  & x\cdot x<0 \\
(\bar x^0;\bar x^i)\,=\,(\eta; r, \th,\phi), \quad \; i,\ j, \ k \mbox{ run from 1 to 3,}&
x\cdot x>0.
\end{cases}
\ee
All quantities in $\bar x^\mu$ coordinate system are denoted by an over bar.
The Ricci curvature (\ref{riccicurv}) reads
\be %
\bar R_{\mu\nu}^\pm
= \begin{cases} 3l^{-2}\,{\rm diag}(1, -\cosh^2(\psi/l),-\cosh^2(\psi/l)
\sin^2\th, 0) & x\cdot x<0, \\
-3l^2\,{\rm diag}(0, -1, -\sinh^2(r/l), -\sinh^2(r/l) \sin^2\th)&
x\cdot x >0.
\end{cases}
\ee

They show that the manifolds are, at least, local $dS_3\times \mathbb{R}$ for
$x\cdot x<0$ and local $\mathbb{R}\times \mathbb{H}_3$ for $x\cdot x
>0$, respectively.
For the former case, $\rho \in (0,\infty)$ or $(-\infty, 0)$, $\psi \in
(-\infty,+\infty)$, $\th \in [0,\pi]$, and $\phi \in [0,2\pi)$. For the latter
case, $\eta \in (0,\infty)$ or $(-\infty,0)$, $r\in [0, \infty)$, $\th \in [0,\pi]$, and
$\phi \in [0,2\pi)$\footnote{The antipodal identification is not taken here
as in \cite{dSSR}.}.  Both 3d \dS\
space-time and 3d hyperboloid have `radius' $l$.

Under the coordinate transformation (\ref{CoordTrans}), the pure Lorentz transformations
$L^\mu_{\ \nu}$ do not induce
singular transformations and not alter $\rho$ and $\eta$.  They will induce the
transformations on $dS_3$ or $\mathbb{H}_3$. However, the points on the
hypersurface satisfying
\be \label{Infinitypts}
1+l^{-1} b\cdot x =0
\ee
are transformed to infinity in the new coordinate system $x'$, meanwhile the infinity
points in the coordinate system $x$, satisfying
\be
(1+l^{-1} b\cdot x)^{-1}=0,
\ee
may be transformed to finite points.  In particular, when $L^\mu_{\ \nu}=\dl^\mu_\nu$ and
$b_\mu \neq 0$ $\forall \mu$,
the points satisfying (\ref{Infinitypts}) correspond to the points transformed from
$(\psi,\th,\phi;\rho)$ to $(\psi, \th, \phi; 0)$ for
$x\cdot x<0$ and from $(\eta;r,\th,\phi)$ to $(0; r,\th,\phi)$ for $x\cdot x>0$.
It implies that the points with $\rho=0$ for $x\cdot x<0$ and with $\eta=0$ for
$x\cdot x>0$
should be in the space-times.  Since the geometries (\ref{g3}), (\ref{h1}), (\ref{connect+})
and (\ref{connect-}) are regular at $\rho=0$ for $x\cdot x<0$ and at $\eta=0$
for $x\cdot x>0$,
the geometries can be extended through $\rho=0$ and $\eta=0$, respectively.
Therefore, $\rho^{-1} \in (-\infty,\infty)$ for $x\cdot x<0$
and $\eta^{-1} \in (-\infty, +\infty)$ for $x\cdot x>0$,
the manifolds
are globally $dS_3\times \mathbb{R}$ for $x\cdot x<0$ and $\mathbb{R} \times \mathbb{H}_3 $
for $x\cdot x>0$, respectively.

Furthermore, the 4d volume elements on the manifolds, defined by
\be
\vect{\eps}= \begin{cases}
l^2\cosh^2(\psi/l) \sin\th d\psi \wedge d\th \wedge d\phi\wedge d\rho,&
{\rm for\ }x\cdot x<0 \\
l^2\sinh^2(r/l) \sin\th d\eta\wedge dr\wedge d\th \wedge d\phi,& {\rm for\ }x\cdot x>0
\end{cases}
\ee
are invariant under the $P_2$ transformation, so the manifolds
are orientable. For $x\cdot x<0$, the manifold is obviously time
orientable because 3d \dS\ space-time is. For $x\cdot x>0$, the
invariant tensor ${\vect h}$ defines an invariant vector field
$\partial_{\eta}$ which is regular on the whole manifold. Compared
with the Newton-Cartan case, it gives an absolute time direction
and, therefore, the space-time is also obviously time orientable.


\section{Uniqueness}

\subsection{Uniqueness of $\mathfrak{p}_2$-invariant metrics}

In the coordinate systems (\ref{Transf2dS}) or
(\ref{Transf2AdS}), $H'$ and $\vect{P}'$ can be written as
\begin{eqnarray}
\left \{ \begin{array}{l}
H'=c \sinh(\psi/l)\d \r {\partial \rho},\\
P'_1=\cosh(\psi/l)\sin\theta\cos\phi\d \r {\partial \rho},\\
P'_2=\cosh(\psi/l)\sin\theta\sin\phi\d \r {\partial \rho},\\
P'_3=\cosh(\psi/l)\cos\theta\d \r {\partial \rho},
\end{array} \right . \qquad  x \cdot x <0, \\
\left \{ \begin{array}{l}
H'=c \cosh(r/l)\d \r {\partial \eta}, \\
P'_1=\sinh(r/l)\sin\theta\cos\phi\d \r {\partial \eta},\\
P'_2=\sinh(r/l)\sin\theta\sin\phi\d \r {\partial \eta},\\
P'_3=\sinh(r/l)\cos\theta\d \r {\partial \eta},
\end{array} \right . \qquad  x\cdot x>0,
\end{eqnarray}
respectively.  Obviously, if the vectors in $\mathfrak{p}_2$-translation
subalgebra $T'$ spanned by $H'$ and $\vect{P}'$ are denoted by
$\vect{\xi}_{(\si)}$, where the subscript in parenthesis $(\si)$
is used to distinguish different vectors, their components have the form of
$\xi_{(\si)}^\la=f_{(\si)}(\psi,\theta,\phi)\dl^\la_3$ for $x\cdot x <0$ or
$\xi_{(\si)}^\la=f_{(\si)}(r,\theta,\phi)\dl^\la_0$ for $x\cdot x > 0$,
respectively.  The direct calculations show that all $\vect{\xi}_{(\si)}$ in Lorentz
algebra $\mathfrak{L}_p$ only
depend on $(\psi, \, \th,\, \phi)$ or $(r,\, \th,\, \phi)$ and $\xi_{(\si)}^{\ 3}=0$
 or $\xi_{(\si)}^{\ 0}=0$ for $x\cdot x \lessgtr 0$, respectively.

\omits{Need to be checked.
\be
\begin{cases}
\vect{J}_1=x_2 \r_3-x_3\r_2 =
\sin\phi\d \r {\r \th}+\cot\th \cos\phi \d \r {\r \phi} &\\
\vect{J}_2=x_3 \r_1-x_1\r_3 =
- \cos\phi\d \r {\r \th}-\cot\th \sin\phi \d \r {\r \phi} &\\
\vect{J}_3= x_1 \r_2-x_2\r_1 =
- \d \r {\r \phi} &\\
\vect{K}_1=x_0 \r_1-x_1\r_0 = l
\sin\th\cos \phi \d \r {\r \psi}+
\tanh (\psi/l) \cos\th  \cos \phi \d \r {\r \th}-
\tanh (\psi/l) \d {\sin\phi}{\sin \th}  \d \r {\r \phi} &\\
\vect{K}_2 =
x_0 \r_2-x_2\r_0 = l
\sin\th\sin\phi\d \r {\r \psi}+
\tanh(\psi/l) \cos\th \sin\phi  \d \r {\r \th}+
\tanh(\psi/l) \d {\cos\phi}{\sin\th}  \d \r {\r \phi} &\\
\vect{K}_3 =
x_0 \r_3-x_3\r_0 = l
 \cos\th \d \r {\r \psi}-\tanh(\psi/l) \sin\th \d \r {\r \th}. 
\end{cases}\label{sphlast}
\ee

\be
\begin{cases}
\vect{J}_1=x_2 \r_3-x_3\r_2 =
\sin\phi\d \r {\r \th}+\cot\th \cos\phi \d \r {\r \phi} &\\
\vect{J}_2=x_3 \r_1-x_1\r_3 =
- \cos\phi\d \r {\r \th}-\cot\th \sin\phi \d \r {\r \phi} &\\
\vect{J}_3= x_1 \r_2-x_2\r_1 =
- \d \r {\r \phi} &\\
\vect{K}_1=x_0 \r_1-x_1\r_0 = l
\sin\th\cos \phi \d \r {\r r}+
\coth (r/l) \cos\th  \cos \phi \d \r {\r \th}-
\coth (r/l) \d {\sin\phi}{\sin \th}  \d \r {\r \phi} &\\
\vect{K}_2 =
x_0 \r_2-x_2\r_0 = l
\sin\th\sin\phi\d \r {\r r}+
\coth(r/l) \cos\th \sin\phi  \d \r {\r \th}+
\coth(r/l) \d {\cos\phi}{\sin\th}  \d \r {\r \phi} &\\
\vect{K}_3 =
x_0 \r_3-x_3\r_0 = l
 \cos\th \d \r {\r r}-\coth(r/l) \sin\th \d \r {\r \th}. 
\end{cases}\label{sphlast}
\ee}

Suppose ${\vect g}^\pm$ are $\mathfrak{p}_2$-invariant
metrics of the covariant form and ${\vect h}_\pm$ are $\mathfrak{p}_2$-invariant
metrics of the contravariant form.  They satisfy
\begin{eqnarray}
&&{\cal L}_{\vect \xi} \bar g^\pm_{\ \mu\nu}=\xi^\la\r_\la \bar g^\pm_{\ \mu\nu} + \bar g^\pm_{\ \mu\la}
\partial_\nu\xi^\la
+\bar  g^\pm _{\ \nu\la}\partial_\mu\xi^\la=0,\qquad\ \forall{\vect \xi}\in\mathfrak{p}_2,
 \label{6A}\\
&&{\cal L}_{\vect \xi}\bar  h_\pm^{\ \mu\nu}=\xi^\la\r_\la\bar  h_{\pm}^{\ \mu\nu} -\bar  h_\pm^{\ \mu\la}
\partial_\la\xi^\nu
- \bar h_\pm ^{\ \nu\la}\partial_\la\xi^\mu=0,\qquad\forall{\vect \xi}\in\mathfrak{p}_2.
\label{6B}
\end{eqnarray}
\omits{(Note: here we don't know whether $g_{ab}$ is degenerate or
not.)}

Consider the case $x \cdot x < 0$ first. Eq. (\ref{6A}) for ${\vect \xi}_{(\si)}\in T'$ reads
\begin{eqnarray}
&&\qquad  0=f_{(\si)}\partial_3\bar g^+_{\ 33}+2\bar g^+_{\ 33}\partial_3f_{(\si)}
=f_{(\si)}\partial_3\bar g^+_{\ 33} \label{d_g33} \\
&& \qquad
0=f_{(\si)}\partial_3\bar g^+_{\ 3\al}+\bar g^+_{\ 33}\partial_\al f_{(\si)}+
\bar g^+_{\ \al 3}
\partial_3f_{(\si)}=
f_{(\si)}\partial_3\bar g^+_{\ 3\al}+\bar g^+_{\ 33}\partial_\al f_{(\si)} \label{g33}\\
&&\qquad 0=f_{(\si)}\partial_3\bar g^+_{\ \al\al}+2\bar g^+_{\ \al 3}\partial_\al
f_{(\si)} \qquad (\mbox{no summation for }\al)\label{g3al}\\
&&\qquad 0 =f_{(\si)}\partial_3\bar g^+_{\ \al\beta}+\bar g^+_{\ \al 3}\partial_\beta
f_{(\si)}+
\bar g^+_{\ \beta 3}\partial_\al f_{(\si)}. \label{d_galbeta}
\end{eqnarray}
Eq.(\ref{d_g33}) gives $\partial_3\bar g^+_{\ 33}=0$ right away.
The validity of Eq.(\ref{g33}) for all $(\si)$ at the same time requires
$\partial_3\bar g^+_{\ 3\al}=\bar g^+_{\ 33}=0$.  Similarly, Eq.(\ref{g3al}) results in
$\partial_3\bar g^+_{\ \al\al}=\bar g^+_{\ \al 3}=0$.  Then, Eq.(\ref{d_galbeta}) leads to
$\partial_3\bar g^+_{\ \al\beta}=0$.
The nontrivial equations of Eq.(\ref{6A}) for ${\vect \xi}_{(\si)}\in\mathfrak{L}_p$ are
\begin{eqnarray}
\xi^\ga\partial_\ga \bar g^+_{\ \al\beta}+\bar g^+_{\ \al \ga}\partial_\beta\xi^\ga+
\bar g^+_{\ \beta\ga}\partial_\al\xi^\ga=0.
\end{eqnarray}
This is nothing but the Killing equation on $dS_3$ on which the 3d metric tensor
$^3{\vect g}$ is unique up to an overall constant scale factor.  Thus, the 4d degenerate
metric in the coordinate system (\ref{Transf2dS}) takes the form
\begin{eqnarray}
(\bar g_{\mu\nu})={\rm diag}(1, -\cosh^2(\psi/l),-\cosh^2(\psi/l)\sin^2\theta,0),
\end{eqnarray}
in which the overall constant scale factor has been chosen as 1.
Similarly, Eq.(\ref{6B}) for ${\vect \xi}_{(\si)}\in T'$ reads
\be
&&0=\omits{\xi^3_{(\si)}\r_3\bar h_{+}^{\ 33} - 2\bar h_+^{\ 3\ga}
\partial_\ga \xi^3_{(\si)} =}f_{(\si)}\r_3\bar h_{+}^{\ 33}- 2\bar h_+^{\ 3\ga}
\partial_\ga f_{(\si)} \label{d_h33}\\
&&0=\omits{\xi^3_{(\si)}\r_3\bar h_{+}^{\ \al 3} -\bar  h_+^{\ \al \ga}
\partial_\ga \xi^3_{(\si)} - \bar h_+^{\ 3\ga}
\partial_\ga \xi^\al_{(\si)}=}f_{(\si)}\r_3\bar h_{+}^{\ \al 3} -\bar  h_+^{\ \al \ga}
\partial_\ga f_{(\si)}\label{h33}\\
&&0=\omits{\xi_{(\si)}^3\r_3\bar h_{+}^{\ \al\beta} -\bar  h_+^{\ \al \ga}
\partial_\ga \xi_{(\si)}^\beta -\bar  h_+ ^{\ \beta \ga}\partial_\ga\xi_{(\si)}^\al =}
f_{(\si)}\r_3\bar h_{+}^{\ \al\beta}.
\end{eqnarray}
They demands that $\r_3\bar h_{+}^{\ 33}=\bar h_+^{\ 3\ga}=\r_3\bar h_{+}^{\ \al 3}
=\bar h_+^{\ \al \beta}
=\r_3\bar h_{+}^{\ \al\beta}=0$.  Eq.(\ref{6B}) for ${\vect \xi}_{(\si)}\in\mathfrak{L}_p$
reads
\be
&&0=\xi^\ga_{(\si)}\r_\ga\bar h_{+}^{\ 33},  \\
&&0=\xi^\ga_{(\si)}\r_\ga\bar h_{+}^{\ \al 3} -\bar  h_+^{\ 3\ga}
\partial_\ga \xi^\al_{(\si)}=\xi^\ga_{(\si)}\r_\ga\bar h_{+}^{\ \al 3},\\
&&0=\xi_{(\si)}^\ga\r_\ga\bar h_{+}^{\ \al\beta} -\bar  h_+^{\ \al \ga}
\partial_\ga \xi_{(\si)}^\beta -\bar  h_+ ^{\ \beta \ga}\partial_\ga\xi_{(\si)}^\al
=\xi_{(\si)}^\ga\r_\ga\bar h_{+}^{\ \al\beta} .
\ee
They constrains $\r_\ga\bar h_{+}^{\ 33}=\r_\ga\bar h_{+}^{\ \al 3}
=\r_\ga\bar h_{+}^{\ \al \beta}=0$.
Therefore, ${\vect h}_+ = - \r_\rho \otimes \r_\rho$ is unique up to a constant scale factor.

Next, consider the case $x \cdot x>0$.  Eq. (\ref{6A}) for ${\vect \xi}_{(\si)} \in T'$ reads
\begin{eqnarray}
&00:&\qquad  0=f_{(\si)}\partial_0\bar g^-_{\ 00}+2\bar g^-_{\ 00}\partial_0f_{(\si)}
=f_{(\si)}\partial_0\bar g^-_{\ 00} \label{d_g00} \\
&0i:& \qquad
0=f_{(\si)}\partial_0\bar g^-_{\ 0i}+\bar g^-_{\ 00}\partial_if_{(\si)}+
\bar g^-_{\ i0}\partial_0f_{(\si)}=
f_{(\si)}\partial_0\bar g^-_{\ 0i}+\bar g^-_{\ 00}\partial_if_{(\si)} \label{g00}\\
&ii:&\qquad 0=f_{(\si)}\partial_0\bar g^-_{\ ii}+2\bar g^-_{\ i0}\partial_if_{(\si)}\label{g0i}\qquad
\mbox{(no summation for $i$)}\\
&ij:&\qquad 0 =f_{(\si)}\partial_0\bar g^-_{\ ij}+\bar g^-_{\ i0}\partial_jf_{(\si)}+
\bar g^-_{\ j0}\partial_if_{(\si)}. \label{d_gij}
\end{eqnarray}
They give rise to $\partial_0\bar g^-_{\ 00}=\partial_0\bar g^-_{\ 0i}=\bar g^-_{\ 00}=
\partial_0\bar g^-_{\ ii}=\bar g^-_{\ i0}=\partial_0\bar g_{ij}=0$.
Eq.(\ref{6A}) for ${\vect \xi}_{(\si)}\in\mathfrak{L}_p$ requires
\begin{eqnarray}
0=\xi^k \partial_k \bar g^-_{\ ij}+\bar g^-_{\ ik} \partial_j\xi^k +
\bar g^-_{\ jk}\partial_i\xi^k,
\end{eqnarray}
which is again just the Killing equation on $\Sigma$. So, $\bar g_{ij}$ is
unique up to a scale factor.  Without loss of generality,
\begin{eqnarray}
(\bar g^-_{\ \mu\nu})={\rm diag}(0, -1, -\sinh^2r, -\sinh^2r\sin^2\theta).
\end{eqnarray}
Similarly, Eq.(\ref{6B}) for ${\vect \xi} \in {\mathfrak p}_2$ sets
$\partial_0\bar h_-^{\ 00}=\bar h_-^{\ 0i}=\partial_0\bar h_-^{\ 0i}=\bar h_-^{\ ik}=
\partial_i \bar h_-^{\ 00}=\partial_i \bar h_-^{\ 0j}=\partial_i \bar h_-^{\ jk}=0$.
Therefore, ${\vect h}_-=\bar h_-^{\ 00}\partial_0\otimes\partial_0$ up to a scale factor.

Therefore, we come to the following theorem.
\begin{thm}(The uniqueness of $\mathfrak{p}_2$-invariant
metrics)\\
Up to an overall constant scale factor,\\
(1) The type-(0,2) degenerate symmetric tensor fields ${\vect g}^\pm$ (\ref{g1})
are unique $\mathfrak{p}_2$-invariant
metrics of the covariant form for $x\cdot x<0$ and $x\cdot x>0$, respectively; and\\
(2) the type-(2,0) degenerate symmetric tensor fields ${\vect h}_\pm$ (\ref{gp2inv}) are unique
$\mathfrak{p}_2$-invariant metrics of the
contravariant form for $x\cdot x<0$ and $x\cdot x>0$, respectively.
\end{thm}

\subsection{The uniqueness of $\mathfrak{p}_2$-invariant connection}
\begin{thm}(The uniqueness of $\mathfrak{p}_2$-invariant
connection)\\
Suppose $\nabla$ is a connection which satisfies\\
1. $\nabla {\vect g}^\pm=0$;\\
2. $\nabla {\vect h}_\pm=0$;\\
3. $[{\cal L}_{\vect \xi},\ \nabla ] {\vect v}=0$, $\forall {\vect \xi} \in\mathfrak{p}_2$
and $\forall {\vect v} \in TM$, \\
then $\nabla$ is unique.
\end{thm}
{\bf Proof:} Taking $x\cdot x>0$ as an example.  In coordinate system
$\bar x^\mu$, the 3-d induced connection is uniquely determined by
$\bar g^-_{\ ij}$ and the unknown
components of connection are $\bar \Ga_{-\eta\eta}^\eta$, $\bar \Ga_{-i\eta}^\eta$,
$\bar \Ga_{-ij}^\eta$, $\bar \Ga_{-\eta\eta}^i$ and $\bar \Ga_{-j\eta}^i$
because of the first condition.  The second condition requires $
\bar \Ga_{-\eta\eta}^\eta=\bar \Ga_{-i\eta}^\eta=0$.
The third condition for ${\vect \xi}_{(\si)} \in T'$ and
${\vect v}=\r_\eta$ is
\begin{eqnarray}
0&=&[{\cal L}_{{\vect \xi}_{(\si)}},\nabla_\mu]v^\nu=
{\cal L}_{{\vect \xi}_{(\si)}}\del_\mu v^\nu-\del_\mu[{\vect \xi}_{(\si)},
\d \r {\partial \eta}]^\nu={\cal L}_{{\vect \xi}_{(\si)}}\del_\mu
(\d \r {\partial \eta})^\nu \nno \\
&=&
f_{(\si)}\d \r {\partial \eta}\Ga_{-\mu\eta}^\nu-\Gamma_{-\mu\eta}^i(\partial_i
f_{(\si)})\dl^\nu_\eta+\Gamma_{-\eta\eta}^\nu(\partial_\mu f_{(\si)}).
\end{eqnarray}
When $\mu=\nu=\eta$, it reads
\begin{eqnarray}
\bar \Ga_{-\eta\eta}^i\partial_i f_{(\si)}=0.
\end{eqnarray}
This is an over-determined set of linear homogeneous equations for $\bar \Ga_{-\eta\eta}^i$,
which has only zero solution, $\bar \Ga_{-\eta\eta}^i=0$.
When $\mu=k$ and $\nu=\eta$, the equation becomes over-determined sets of linear
homogeneous equations for $\bar \Ga_{-k\eta}^i$:
\begin{eqnarray}
\bar \Ga_{-k\eta}^i\partial_i f_{(\si)}=0,
\end{eqnarray}
which have only zero solutions, $\bar \Ga_{-k\eta}^i=0$, too.
The third condition for ${\vect \xi} \in T'$ and
${\vect v}=\r_i$ is
\begin{eqnarray}
0 &=&[{\cal L}_{{\vect \xi}_{(\si)}},\nabla_\mu]v^\nu
={\cal L}_{{\vect \xi}_{(\si)}}\del_\mu v^\nu -\del_\mu[{\vect \xi}_{(\si)},
\partial_i]^\nu\nonumber\\
&=&f_{(\si)}\d \r {\partial \eta} \bar \Ga_{-\mu i}^\nu-\bar \Ga_{-\mu i}^k(\partial_kf_{(\si)})
\dl^\nu_\eta
+\bar \Ga_{-\eta i}^\nu (\partial_\mu f_{(\si)})+(\partial_\mu\partial_if_{(\si)})
\dl^\nu_\eta +(\partial_i
f_{(\si)})\bar \Ga_{-\mu\eta }^\nu.
\end{eqnarray}
When $\mu=j$, $\nu=\eta $, it reads
\begin{eqnarray}
f_{(\si)}\d \r {\partial \eta} \bar \Ga_{-ji}^\eta -\bar \Ga_{-ji}^k(\partial_kf_{(\si)})
+\partial_j\partial_if_{(\si)} =0, \nonumber
\ee
which leads to
\be
\d \r {\partial \eta} \bar \Ga_{-ji}^\eta =\bar \Ga_{-ji}^k(\partial_k\ln
f_{(\si)})-\frac{\partial_j\partial_if_{(\si)}}{f_{(\si)}}=+g^-_{ij}.
\end{eqnarray}
Thus,
\be
\bar \Ga_{-ji}^\eta = \eta g^-_{ij}+\gamma_{ij}^0
\ee
where $\gamma$ is independent on $\eta$.  The third condition for ${\vect \xi} \in \mathfrak{L}_p$
 and
${\vect v}=\r_i$ is
\begin{eqnarray}
0&=&[{\cal L}_{{\vect \xi}_{(\si)}},\del_\mu ]v^\nu
={\cal L}_{{\vect \xi}_{(\si)}}(\del_\mu v^\nu)-\del_\mu
[{\vect \xi}_{(\si)},\partial_i]^\nu\nonumber\\
&=&\xi_{(\si)}^{\ \ k}\partial_k\Ga_{\mu i}^\nu-\Ga_{\mu i}^\la\partial_\la\xi_{(\si)}^{\ \ \nu}
+\Ga_{\la i}^\nu\partial_\mu\xi_{(\si)}^{\ \ \la}
+\partial_\mu\partial_i\xi_{(\si)}^{\ \ \nu}+\Ga_{\mu k}^\nu\partial_i\xi_{(\si)}^{\ \ k}.
\end{eqnarray}
When $\mu=j$ and $\nu=\eta$, it reads
\begin{eqnarray}
0=\xi_{(\si)}^{\ \ k}\partial_k\bar \Ga_{-ji}^\eta+\bar \Ga_{-ki}^\eta \partial_j\xi_{(\si)}^{\ \ k}
+\bar \Ga_{-jk}^\eta\partial_i\xi_{(\si)}^{\ \ k}
\end{eqnarray}
Since $\eta g^-_{ij}$ satisfies the equation, $\gamma_{ij}^0$ should also satisfies the
equation. It is just the Killing equation if
$\gamma_{ij}^0$ acts as a $(0,2)$-type tensor. \omits{$=C\cdot
g_{ij}$,} Thus, the general form of $\bar \Ga_{-ij}^\eta$ should be
\begin{eqnarray}
\bar \Ga_{-ij}^\eta=(\eta + C)g^-_{ij}.
\end{eqnarray}
It differs from Eq.(\ref{connect-}) trivially by a
simple coordinate transformation $\eta\to \eta + C$, which corresponds to the coordinate
transformation Eq.(\ref{CoordTrans}) with Eq.(\ref{Infinitypts}) and
$L^\mu_{\ \nu}=\dl^\mu_\nu$.\hfill $\Box$


\section{Symmetries}
\subsection{Maximum symmetry of the geometries}
In the above section, we have shown that the geometries $(M^{{\frak p}_2}, {\vect g}^\pm,
{\vect h}_\pm, \del^\pm)$ are the unique geometries which are invariant under the $P_2$
transformation.  In this subsection, we shall show that the Killing vector field ${\vect \xi}$
satisfying Eqs.(\ref{eq:Lieg}),
(\ref{eq:Lieh}) and (\ref{eq:LieGamma}) simultaneously must belong to ${\frak p}_2$.

Now, suppose ${\vect v}$ be an arbitrary vector field. Eq.(\ref{eq:LieGamma}) acting
on $v^\nu$ gives
\be
[{\cal L}_{\vect \xi}, \del^\pm_\mu]v^\nu&=&
\xi^\la (\del^\pm_\la \del^\pm_\mu v^\nu- \del^\pm_\mu \del^\pm_\la v^\nu) +v^\la\del^\pm_\mu
\del^\pm_\la\xi^\nu \nno \\
&=&-\xi^\la R^{\nu}_{\pm \ka\la\mu}v^\ka+v^\la\del^\pm_\mu\del^\pm_\la\xi^\nu=0,
\ee
which implies
\be
\del^\pm_\mu\del^\pm_\la\xi^\nu=R^\nu_{\pm \la \ka\mu}\xi^\ka=
\pm l^{-2}(\dl^\nu_\mu g^\pm_{\ \la\ka}\xi^\ka-g^\pm_{\ \la\mu}\xi^\nu) .
\ee
On the other hand,
\be
\del^\pm_\mu \del^\pm_\la \xi^\nu 
&=&\r_\mu  (\r_\la \xi^\nu +\Ga^\nu _{\pm \la \ka}\xi^\ka )-
\Ga^\ka _{\pm \mu\la}(\r_\ka  \xi^\nu+\Ga^\nu _{\pm \ka \si}\xi^\si)
+\Ga^\nu _{\pm \mu \ka}(\r_\la \xi^\ka +\Ga^\ka _{\pm \la \si}\xi^\si) \nno  \\
\omits{
&=&\r_\mu  \r_\la  \xi^\nu  + \Ga^\nu _{\pm dc}\r_\mu \xi^\ka  +\Ga^\nu _{\pm ac}\r_\la \xi^\ka -\Ga^\ka _{\pm ad}\r_\ka \xi^\nu
+ (\r_\mu  \Ga^\nu _{\pm de}
-\Ga^\ka _{\pm ad}\Ga^\nu _{\pm ce}
+\Ga^\nu _{\pm ac}\Ga^\ka _{\pm de})\xi^e \nno \\
&=&\r_\mu  \r_\la  \xi^\nu  -\d {x_\la \dl^\nu _\ka +x_\ka \dl^\nu _\la }{x\cdot x}\r_\mu \xi^\ka
-\d {x_\mu \dl^\nu _\ka +x_\ka \dl^\nu _\mu }{x\cdot x}\r_\la \xi^\ka  + \d {x_\mu \dl^\ka _\la +x_\la \dl^\ka _\mu }{x\cdot x}
\r_\ka \xi^\nu  \nno \\
&&+ (\r_\mu  \Ga^\nu _{\pm de}
-\Ga^\ka _{\ ad}\Ga^\nu _{\pm ce}
+\Ga^\nu _{\pm ac}\Ga^\ka _{\pm de})\xi^e \nno \\
&=&\r_\mu  \r_\la  \xi^\nu -\d {x_\la \r_\mu \xi^\nu +x_\ka \dl^\nu _\la \r_\mu \xi^\ka }{x\cdot x}
-\d {x_\mu \r_\la \xi^\nu +x_\ka \dl^\nu _\mu \r_\la \xi^\ka }{x\cdot x}+ \d {x_\mu \r_\la \xi^\nu
+x_\la \r_\mu \xi^\nu }{x\cdot x} \nno \\
&&- \left ( \d {\eta_{ad}\dl^\nu _e+\eta_{ae}\dl^\nu _\la }{x\cdot x}
- \d {2x_\mu (x_\la \dl^\nu _e+x_e\dl^\nu _\la )}{(x\cdot x)^2}
+ \d {x_\la \dl^\ka _\mu +x_\mu \dl^\ka _\la }{x\cdot x}\d {x_e\dl^\nu _\ka +x_\ka \dl^\nu _e}{x\cdot x} \right . \nno \\
&& \left . -\d {x_\ka \dl^\nu _\mu +x_\mu \dl^\nu _\ka }{x\cdot x}\d {x_e\dl^\ka _\la +x_\la \dl^\ka _e}{x\cdot x}
\right )\xi^e \nno \\
&=&\r_\mu  \r_\la  \xi^\nu -\d {x_\ka \dl^\nu _\la \r_\mu \xi^\ka +x_\ka \dl^\nu _\mu \r_\la \xi^\ka }{x\cdot x}
-\d {\eta_{ad}\dl^\nu _e+\eta_{ae}\dl^\nu _\la }{x\cdot x}\xi^e
+ \d {2x_\mu (x_\la \dl^\nu _e+x_e\dl^\nu _\la )}{(x\cdot x)^2}\xi^e
\nno \\
&&-\d {x_\la x_e\dl^\nu _\mu +2 x_\la  x_\mu \dl^\nu _e+x_ex_\mu \dl^\nu _\la }{(x\cdot x)^2} \xi^e +
\d {2 x_ex_\la \dl^\nu _\mu +x_ex_\mu \dl^\nu _\la +x_\la x_\mu \dl^\nu _e}{(x\cdot x)^2}
\xi^e \nno \\ }
&=&\r_\mu  \r_\la  \xi^\nu -\d {x_\ka \dl^\nu _\la \r_\mu \xi^\ka +x_\ka \dl^\nu _\mu \r_\la \xi^\ka }{x\cdot x}
\pm l^{-2}g^\pm_{\mu \la}\xi^\nu \pm l^{-2}\dl^\nu _\la  g^\pm_{\mu \si}\xi^\si
+ \d {(x_\mu \dl^\nu _\la +x_\la \dl^\nu _\mu )x_\si}{(x\cdot x)^2}\xi^\si.
\ee
They give rise to the PDE
\be \label{PDE}
\r_\mu  \r_\la  \xi^\nu -\d {x_\ka \dl^\nu _\la \r_\mu \xi^\ka +x_\ka \dl^\nu _\mu \r_\la \xi^\ka }{x\cdot x}
\pm l^{-2}(\dl^\nu _\mu  g^\pm_{\la\si}+\dl^\nu _\la  g^\pm_{\mu\si})\xi^\si
+ \d {x_\mu \dl^\nu _\la +x_\la \dl^\nu _\mu }{(x\cdot x)^2}x_\si\xi^\si=0.
\ee
Multiplication with $x^\mu x^\la x_\nu$, it reduces to
\be
\omits{x^ax^dx_b\r_\mu  \r_\la  \xi^\nu - 2x_\ka x^a\r_\mu \xi^\ka  + 2x_e\xi^e =0.
\ee
It is equivalent to
\be
&& x^ax^d\r_\mu  (x_b\r_\la  \xi^\nu )-x_b x^d \r_\la  \xi^\nu - 2x^a\r_\mu (x_\ka \xi^\ka )+
4x_e\xi^e =0 \nno \\
{\rm or} &&x^ax^d\r_\mu  \r_\la  (x_b\xi^\nu )-2 x^ax_b\r_\mu  \xi^\nu - 2x^a\r_\mu (x_\ka \xi^\ka )+
4x_e\xi^e =0 \nno \\
{\rm or} &&}
x^\mu\r_\mu  (x^\la\r_\la  (x_\nu\xi^\nu ))- 5x^\mu\r_\mu (x_\ka \xi^\ka )+
6x_\ka\xi^\ka =0. 
\ee
It has the following general solution
\be
x_\ka \xi^\ka  = C_1({x^i}/{x^0})
(\pm x\cdot x) +C_2({x^i}/{x^0}) (\pm x\cdot x)^{3/2},
\ee
where $C_1$ and $C_2$ are the functions of the ratio of $x^i$ to $x^0$ to be determined.
Therefore, the Killing vector field should have the form
\be
\xi^\ka  =\pm C_1({x^i}/{x^0}) x^\ka \pm C_2({x^i}/{x^0})\cdot (\pm x\cdot x)^{1/2}x^\ka
+C_3(x)(x_\la\dl^\ka_\si -x_\si\dl^\ka_\la)
+C^\mu(x) g^\pm_{\mu\nu}\eta^{\nu\ka},
\ee
where $C_3(x)$ and $C_\mu(x)$ are the functions of $x$ to be determined.

In order to fix $C_1,\ C_2,\ C_3$ and $C^\mu$, we study the above expression
term by term.  The PDE (\ref{PDE}) for the first term reads
\be
\omits{&&\r_a \r_d (C_1x^b)-\d {x_c\dl^b_d\r_a(C_1x^c)+x_c\dl^b_a\r_d(C_1x^c)}{x\cdot x}
\pm l^{-2}(\dl^b_e g_{de}+\dl^b_d g_{ae})C_1x^e
+ \d {x_a\dl^b_d+x_d\dl^b_a}{(x\cdot x)^2}x_eC_1x^e=0 \nno \\
&&\r_a \r_d (C_1x^b)-\d {x_c\dl^b_dx^c\r_a C_1+x_c\dl^b_ax^c\r_d C_1+x_a\dl^b_d C_1
+x_d\dl^b_a C_1}{x\cdot x}
+ \d {x_a\dl^b_d+x_d\dl^b_a}{x\cdot x}C_1=0 \nno \\
&&\r_a (x^b\r_d C_1+C_1\dl_d^b)-(\dl^b_d\r_a C_1+\dl^b_a\r_d C_1)=0 \nno \\
&&x^b \r_a \r_d C_1 +\dl^b_a\r_dC_1+ \r_a C_1\dl_d^b-(\dl^b_d\r_a C_1+\dl^b_a\r_d C_1)
=0
\nno \\
&& }
\r_\mu \r_\nu C_1 = 0. \omits{\quad \Rightarrow \quad  \r_d C_1 = const.
\quad \Rightarrow \quad}
\ee
Thus, $C_1$ is, at most, the linear function of $x$.  However, $C_1$ is independent of
$x\cdot x$.  Therefore, $C_1$ can only be a non-zero constant, at most.  \omits{In other words, the vector field $\xi=x^\mu\r_\mu$ can preserve the
form of the connection.}
Note that
\be
{\cal L}_{x^\mu\r_\mu} h_\pm^{ab}&=&l^{-4}{\cal L}_{x^\mu\r_\mu}[(x\cdot x)x^\la
x^\si (\r_\la)^a (\r_\si)^b] \omits{\nno \\
&=&l^{-2}{x^\mu\r_\mu}[(x\cdot x)x^\la
x^\si (\r_\la)^a (\r_\si)^b]-2l^{-2}[(x\cdot x)x^\la
x^\si  (\r_\si)^b]\r_\la x^\mu (\r_\mu)^a\nno \\
&=&l^{-2}[{x^\mu\r_\mu}(x\cdot x)]x^\la
x^\si (\r_\la)^a (\r_\si)^b+l^{-2}(x\cdot x){x^\mu\r_\mu}[x^\la
x^\si (\r_\la)^a (\r_\si)^b] \nno \\
&&-2l^{-2}[(x\cdot x)x^\la
x^\si  (\r_\si)^b]\dl_\la^\mu (\r_\mu)^a\nno \\
&=&2 l^{-2}x^\mu x_\mu x^\la
x^\si (\r_\la)^a (\r_\si)^b+l^{-2}(x\cdot x)x^\mu (\dl^\la_\mu x^\si+\dl^\si_\mu x^\la
) (\r_\la)^a (\r_\si)^b]\nno \\
&=&4 h^{ab}-2h^{ab}}=2h_\pm^{ab}\neq 0.
\ee
Therefore, $C_1$ must be zero.

The PDE (\ref{PDE}) for the second term reads
\be
\omits{&&\r_a \r_d ( (\pm x\cdot x)^{1/2}x^bC_2)-\d {x_c\dl^b_d\r_a((\pm x\cdot x)^{1/2}
 x^cC_2)+
x_c\dl^b_a\r_d((\pm x\cdot x)^{1/2}C_2 x^c)}{x\cdot x}
\pm \d {x_a\dl^b_d+x_d\dl^b_a}{(\pm x\cdot x)^{1/2}}C_2 =0 \nno \\
&&\r_a \r_d ((\pm x\cdot x)^{1/2}x^b C_2 )- (\dl^b_d\r_a((\pm x\cdot x)^{1/2}C_2 )+
\dl^b_a\r_d((\pm x\cdot x)^{1/2}C_2 ))=0 \nno \\
&&}
\r_\mu \r_\nu ((\pm x\cdot x)^{1/2}C_2 )=0
\omits{\quad \Rightarrow \quad
 \r_d (C_2 (-x\cdot x)^{1/2})=const.}
\ee
It means that $(\pm x\cdot x)^{1/2}C_2 $ is a linear function of $x$, at most.
Since $C_2$ does not contain the factor $(\pm x\cdot x)^{1/2}$.  The above result implies
that $C_2$ should be the homogeneous linear function of $x^\ka/(\pm x\cdot x)^{1/2}$.
Therefore, the possible linear-independent vector fields $\xi^c_{(\nu)}=x_\nu x^\ka\r_\ka$,
which are proportional to the pseudo-translation generators in ${\frak p}_2$.

The PDE (\ref{PDE}) for the fourth term reduces to
\be
\omits{&&\r_a \r_d (C^e(x) g^\pm_{ec}\eta^{cb})-\d {x_c\dl^b_d\r_a(C^e(x) g^\pm_{ef}\eta^{fc})
+x_c\dl^b_a\r_d(C^e(x) g^\pm_{ef}\eta^{fc})}{x\cdot x} \nno \\
&&\pm l^{-2}(\dl^b_d g^\pm_{ae}+\dl^b_a g^\pm_{de})(C^g(x) g^\pm_{gf}\eta^{fe})
+ \d {x_a\dl^b_d+x_d\dl^b_a}{(x\cdot x)^2}x_e(C^g(x) g^\pm_{gf}\eta^{fe})=0 \nno \\
&&\r_a \r_d (C^e(x) g^\pm_{ec}\eta^{cb})-\d {x^f\dl^b_d\r_a g^\pm_{ef}
+x^f\dl^b_a\r_d g^\pm_{ef}}{x\cdot x} C^e(x)-\d {\dl^b_d g^\pm_{ag}
+\dl^b_a g^\pm_{dg}}{x\cdot x}C^g(x)
=0 \nno \\
&&}\r_\mu \r_\nu (C^\la(x) g^\pm_{\la\ka}\eta^{\ka\si})=0, \omits{\quad \Rightarrow \quad
\r_d (C^e(x) g_{ec}) = const.\quad \Rightarrow \quad
C^e(x) g_{ec}  = A_c x^d +B_c^d}
\ee
which implies
\be
C^\la(x) g^\pm_{\la\ka}  = A_{\ka \nu} x^\nu +B_\ka,
\ee
where $A_{\ka\nu}$ and $B_\ka$ are constants. Since
\be
A_{\ka\nu} x^\ka x^\nu +B_\ka x^\ka =0 \qquad  \forall x,
\ee
there is no nonzero solution for $A_{\ka\nu}$ and $B_\ka$.
It implies
\be
C^\ka(x) g^\pm_{\ka \nu}  = 0.
\ee

Finally, Eq. (\ref{eq:Lieg}) for the third requires
\be
 C_{3,\nu}(x) g^\pm_{\mu[\si}x_{\la]}  +
 C_{3,\mu}(x) g^\pm_{\nu[\si}x_{\la]}=0
\ee
because $x_\la \r_\si-x_\si \r_\la$ are Killing vectors.
When $\mu=\nu$, it reduces to
\be
  C_{3,\mu}(x) g^\pm_{\nu[\si}x_{\la]}=0.
\ee
Since $g^\pm_{\nu[\si}x_{\la]}$ does not always vanish, $C_3$ must be a constant.
\omits{Eq. (\ref{PDE}) for the third term can be written as
\be
\omits{&&\r_a \r_d (C_3(x)(x_\la\dl^b_\si -x_\si\dl^b_\la))
-\d {x_c\dl^b_d\r_a(C_3(x)(x_\la\dl^c_\si -x_\si\dl^c_\la))+x_c\dl^b_a\r_d
(C_3(x)(x_\la\dl^c_\si -x_\si\dl^c_\la))}{x\cdot x} \nno \\
&&\pm l^{-2}C_3(x)(\dl^b_d g^\pm_{ae}+\dl^b_a g^\pm_{de})(x_\la\dl^e_\si -x_\si\dl^e_\la)
=0 \nno \\
&&\r_a \r_d (C_3(x)(x_\la\dl^b_\si -x_\si\dl^b_\la))
-\d {\dl^b_d(\eta_{a\la}x_\si -\eta_{a\si}x_\la)+\dl^b_a
(\eta_{d\la}x_\si -\eta_{d\si}x_\la)}{x\cdot x}C_3(x) \nno \\
&&\pm l^{-2}[\dl^b_d (x_\la g^\pm_{a\si} -x_\si g^\pm_{a\la})
+\dl^b_a (x_\la g^\pm_{d\si} -x_\si g^\pm_{d\la})]C_3(x)
=0\nno \\
&&\r_a \r_d (C_3(x)(x_\la\dl^b_\si -x_\si\dl^b_\la))=0
\quad \Rightarrow \quad \r_d (C_3(x)(x_\la\dl^b_\si -x_\si\dl^b_\la)) = const.\nno \\
&&}(x_\la\dl^b_\si -x_\si\dl^b_\la)\r_d C_3(x)+C_3(x)
(\eta_{d\la}\dl^b_\si -\eta_{d\si}\dl^b_\la) = const.
\ee
It reduces to
\be
(x_\la\r_\si -x_\si\r_\la)C_3(x) = const,
\ee
after the summation for $b=d$ is taken.
\omits{\be
&&(x_\la\r_\si -x_\si\r_\la)(x^\la\r^\si -x^\si\r^\la)C_3(x) = 0 \nno \\
&{\rm or\ }&2((x\cdot x) \eta^{\la\si}\r_\la\r_\si-x^\la\r_\la x^\si\r_\si -2x^\la\r_\la)C_3=0 \nno\\
&{\rm or\ }&(\eta^{\la\si}\r_\la\r_\si +\r_\rho^2+\d 3 \rho \r_\rho)C_3=0\nno \\
&{\rm or \ }&[(\eta^{ij} + \frac {x^i}{x^0} \frac {x^j}{x^0})\r_{x^i/x^0}\r_{x^j/x^0}
+2 \frac {x^i}{x^0}\r_{x^i/x^0}]C_3=0 \nno \\
&{\rm or\ }&\rho^{-2}\{\r_\psi^2+2\tanh\psi \r_\psi -\d 1 {\cosh^2\psi}[\d {1}{\sin \th}
\r_\th (\sin\th \r_\th)+\d 1 {\sin^2\th}\r_\phi^2]\}C_3=0
\ee}
It implies that $C_3(x)=C_3(x^i/x^0)$, at most.

\omits{Its solution has the form
\be
C_3=f(\rho)\bar C_3(\psi, \th, \phi)=f(x\cdot x)\sum_{l,m}\Psi_l(\psi)Y_{lm}(\th,\phi),
\ee
where $Y_{lm}(\th,\phi)$ is the spherical harmonic functions and $\Psi_l$ satisfies
\be
(\cosh^2\psi\Psi'_l)' +l(l+1)\Psi_l=0.
\ee
It can be rewritten as
\be
\d {d^2\Psi_l}{d \xi^2} +\d {l(l+1)}{1-\xi^2}\Psi_l=0.
\ee
where $\xi=\tanh\psi$.  Its solution is hypergeometric function
\be
\Psi_l=A_l F(-\d {1+l}{2},\d l 2, \d 1 2;\tanh^2\psi)+B_l\tanh\psi F
(\d {1+2}2,-\d {l} 2,\d{3}2;\tanh^2\psi).
\ee
In particular, when $l=0$,
\be
\Psi_0=A_0 +B_0\tanh\psi .
\ee}

Substituting the solution in the original equation, we have
\be
(x_\la\dl^b_\si -x_\si\dl^b_\la)\r_d (f\bar C_3)+f\bar C_3
(\eta_{d\la}\dl^b_\si -\eta_{d\si}\dl^b_\la) = C^b_{d\la\si}=const.
\ee
For $\la\neq \si =b=0$, it reduces to
\be
 x_\la \r_d (f\bar C_3)+f\bar C_3 \eta_{d\la} = C^0_{d\la 0}=const.
\ee
Multiplying $\eta^{d\la}$, we have
\be
(\rho \d {df}{d\rho}+4f)\bar C_3= C^0_{d\la 0}\eta^{d\la} =const.
\ee
It requires that if the constant $C^0_{d\la 0}\eta^{d\la}\neq 0$,
$\bar C_3$ is a constant and
\be
\rho\d {df}{d\rho}+4f= 4a=C^0_{d\la 0}\eta^{d\la}/\bar C_3,
\ee
and that if the constant $C^0_{d\la 0}\eta^{d\la}= 0$,
\be
\rho\r_\rho f+4f=0
\ee
while $\bar C_3$ has no further constraint.
They have the solutions
\be
f=\begin{cases} a+\d b {\rho^4} & \mbox{for } C^0_{d\la 0}\eta^{d\la}\neq 0\\
\d b {\rho^4} &\mbox{for } C^0_{d\la 0}\eta^{d\la}= 0.
\end{cases}
\ee
Further, for $d=0$, $\la=3$, we have
\be
&& \begin{cases} x_3 \bar C_3 \r_0 f = C^0_{030}=const.&
\mbox{for } C^0_{d\la 0}\eta^{d\la}\neq 0 \\
  x_3 \bar C_3 \r_0 f +x_3 f\r_0\bar C_3= C^0_{030}=const. &
  \mbox{for } C^0_{d\la 0}\eta^{d\la}= 0.\end{cases}\\
&{\rm i.e.} & \begin{cases}
4b \rho^{-4}\sinh\psi\cosh \psi \cos\th
=C^0_{030}/\bar C_3 =const. & \mbox{for } C^0_{d\la 0}\eta^{d\la}\neq 0 \\
4b \rho^{-4}\sinh\psi\cosh \psi \cos\th \bar C_3
+ b\rho^{-4}\cosh^2 \psi \cos\th
 \r_\psi \bar C_3 =C^0_{030} &  \mbox{for } C^0_{d\la 0}\eta^{d\la}= 0\end{cases}  \nno \\
&{\rm i.e.} &\begin{cases}
b=0,\qquad  C^0_{030}=0 & \\
\d {b\cos\th} {\rho^4}  \left (\d {4\xi} {1-\xi^2} \bar C_3
+ \r_\xi \bar C_3 \right )=C^0_{030}& \end{cases}
\ee
The latter equation requires that $C^0_{030}$ and either $b=0$ or
\be
\d {4\xi} {1-\xi^2} \Psi_l + \r_\xi \Psi_l =0,
\ee
which contradicts to the equation for $\Psi_l$.  Therefore, the non-zero solution for
$C_3$ must be a non-vanishing constant, which can be taken 1 without loss of generality.
($C^0_{030}=b=0$ but
$C^0_{d\la 0}\eta^{d\la}\neq 0$.)
Thus, the possible vector fields have the form $\xi_{(\si)(\la)}=x_\si\r_\la -x_\la \r_\si$.
}

Then, we come to the theorem.
\begin{thm} The maximum symmetry of the geometries $(M^{{\frak p}_2}, {\vect g}^\pm,
{\vect h}_\pm, \del^\pm)$ with Eqs.(\ref{g1}), (\ref{gp2inv}) and (\ref{inv-connectn})
is the second Poincar\'e group.
\end{thm}

\subsection{Re-classification of the symmetry}
In the algebraic point of view, the so-called pseudo-translation
generators, $H'$ and $\vect{P}'$, spanning the Abelean ideal of $\frak{iso}(1,3)$,
take the role of the time and
space translation ones, respectively,
and $\vect{K}$ and $\vect{J}$ span the ${\frak{so}}(1,3)$
algebras as usual, generating the $SO(1,3)$ isotropy of
space-time.  Its subalgebra ${\frak so}(3)$ generates the $SO(3)$
isotropy of space. However,  the decomposition does not fit the
above structure of space-time.

For the case $x\cdot x<0$, the manifold is $dS_3\times \mathbb{R}$.
The metric of the $dS_3$ space-time can be written as
\be
ds^2=\d {\eta_{\al\beta}d z^\al dz^\beta}{\si_3(z)}
+\d {(\eta_{\al\beta}z^\al dz^\beta)^2}{l^2\si_3^2(z)}
\ee
in terms of a 3d Beltrami coordinate system, say, on the chart $U_3$ \cite{dSSR},
\be
z^0 &=& l\d {x^0}{x^3},\quad  z^1=l\d {x^1}{x^3}, \quad  z^2=l\d {x^2}{x^3},
\ee
where
\be
\si_3(z)= 1-l^{-2} \eta_{\al \beta}z^\al z^\beta >0,
\ee
and $\al,\; \beta$ run over 0, 1, 2. On the $dS_3$ space-time,
there are 3d Beltrami translations,
which take the role of translation in the
neighborhood of the origin on the 3d manifolds.  They are
\be
\begin{cases}
^3H^+ = c \r_{z^0}-cl^{-2}z_0 z^\beta\r_{z^\beta}\omits{= \d {c^2} {l} K_3}=:{\cal H}, & \\
{}^3P^+_1 = \r_{z^1}-l^{-2}z_1 z^\beta \r_{z^\beta}\omits{=- l^{-1}J_2}
=:{\cal P}_1, & \\
{}^3P^+_2 = \r_{z^2}-l^{-2}z_2 z^\beta\r_{z^\beta} \omits{=l^{-1}J_1}
=:{\cal P}_2,&
\end{cases}
\ee
where the superscript 3 stands for the quantity being defined on the 3d space-time.
The pseudo-translation generator in $x^3$ defines the translation in direction
$z^3=l^2/x^3 =\rho {\rm sech}(\psi/l) {\sec\th}$
\be \label{P3'+}
P'_3=\r_{z^3} =:{\cal P}_3.
\ee
The boost generators in 3d dS space-time are
\be
\begin{cases}
{}^3{K}_1=\d 1 c (z_0\r_{z^1}-z_1\r_{z^0})={K}_1=:{\cal K}_1, & \\
{}^3{K}_2
=\d 1 c(z_0\r_{z^2}-z_2\r_{z^0})={K}_2=:{\cal K}_2. &
\end{cases}
\ee
The Galilei boost in the direction $z^3$ is
\be \label{H'+}
\d 1 c z_0 \r_{z^3}=\d l {c^2}  H'=:{\cal K}_3.
\ee
The three space `rotation' generators
\be \label{J+}
\begin{cases}
{{\cal J}}_1:= {z_1}\r_{z^3}=l{P}'_1, & \\
{{\cal J}}_2:=
{z_2}\r_{z^3}=l{P}'_2, & \\
{{\cal J}}_3:=z_1\r_{z^2}-z_2\r_{z^1}={J}_3, &
\end{cases}
\ee%
spans an $\frak{iso}(2)$ subalgebra, %
\be%
[{{\cal
J}}_1,{{\cal J}}_2]=0,\quad  [{{\cal J}}_1, {{\cal
J}}_3]=-{{\cal J}}_2,\quad [{{\cal J}}_2,{{\cal
J}}_3]={{\cal J}}_1. %
\ee
Finally, it can be shown that
\begin{eqnarray}
\begin{cases}
\frac {c^2} l K_3 = {\cal H} - \frac {c^2}
{l^2} z^33{\cal K}_3 \\
-l^{-1}J_2 = {\cal P}_1 - l^{-2}z^3{\cal J}_1\\
l^{-1}J_1 = {\cal P}_2 - l^{-2}z^3{\cal J}_2
\end{cases} \qquad {\rm or} \qquad
\begin{cases}
\frac {c^2} l K_3 + \frac l {x^3}H' = {\cal H} \\
l^{-1}J_2 + \frac l {x^3} P'_1 = {\cal P}_1 \\
l^{-1}1J_1 + \frac l {x^3} P'_2 = {\cal P}_2
\end{cases}.
\end{eqnarray}

The set of generators $({\cal H}- (c^2/l^2)z^3{\cal K}_3,{\cal P}_1 - l^{-2}z^3{\cal J}_1,
{\cal P}_2 - l^{-2}z^3{\cal J}_2, {\cal P}_3, {\vect{\cal K}},
{\vect{\cal J}})$ defines an alternative decomposition of
${\frak{iso}}(1,3)$ algebra, different from the Poincar\'e algebra.
Clearly, the alternative decomposition fits to the geometrical
structure.  The decomposition together the geometrical structure
gives a new realization of ${\frak{iso}}(1,3)$ algebra.  It defines
a new possible kinematics without the space SO(3) isotropy.

For the case $x\cdot x>0$, the manifold is $\mathbb{R}\times \mathbb{H}_3$.
In terms of the Beltrami coordinates, the metric of $\mathbb{H}_3$ space is
\be
ds^2=-\d {\dl_{ij} dz^i dz^j}{\si_3^E(z)}
-\d {(\dl_{ij}z^i dz^j)^2}{(l\si_3^{E}(z))^2},
\ee
where
\be\label{z}
z^i &=&l\d {x^i}{x^0},
\ee
and
\be
\si_3^{E}(z)=1-l^{-2}\dl_{ij}z^i z^j>0.
\ee
On $\mathbb{H}_3$ space, the 3d Beltrami translations
\be
{}^3{P}^-_i &=& \r_{z^i}+l^{-2}z_i z^j\r_{z^j}=\d c l {K}_i=:\tilde{{\cal P}}_i,
\ee
play the role of translation in the
neighborhood of the origin on the 3d manifold, and the space rotation generators,
defined by
\be
\tilde {{\cal J}}_i:=\d 1 2 \eps_{i}^{\ jk}(z_j{\cal P}_k
- z_k{\cal P}_j )={J}_i,
\ee
span the $\frak{so}(3)$ subalgebra
\be
[\tilde{{\cal J}}_i,\tilde{{\cal J}}_j]=
-\eps_{ij}^{\ \ k}\tilde{{\cal J}}_k.
\ee
Define $z^0=l^2/x^0= \eta {\rm sech} (r/l)$.  Then, the Carroll boosts
\be \label{P'-}
\tilde{{\cal K}}_i:=\d 1 c {z_i}\r_{z^0}=
\d l c {P}'_i.
\ee
The pseudo-time translation
\be \label{H'-}
H'=  c \r_{z^0}=:\tilde{\cal H}
\ee
defines the time translation in $z^0$ direction.
Finally, it can be shown that
\begin{eqnarray}
\dfrac c  l K_i = \tilde {\cal P}_i +
\dfrac c {l^2} z^0\tilde {\cal K}_i \qquad {\rm or} \qquad
\frac c l K_i - \dfrac l {x^0} \tilde {\cal P}_i.
\end{eqnarray}
$(\tilde{\cal H}, \tilde{\vect{\cal P}}+ (c/l^2)z^0
\tilde{\vect{\cal K}}, \tilde{\vect{\cal K}},
\tilde{\vect{\cal J}})$
gives another alternative decomposition of $\frak{iso}(1,3)$ algebra,
\begin{eqnarray}
\begin{cases}
[\tilde{\cal H}, \tilde{{\cal P}}_i+\dfrac c {l^2}z^0\tilde{{\cal K}}_i]
= \d {c^2} {l^2} \tilde{{\cal K}}_i,
\quad  [\tilde{{\cal P}}_i +\dfrac c {l^2}z^0\tilde{{\cal K}}_i,
\tilde{{\cal P}}_j + \dfrac c {l^2} z^0 \tilde{{\cal K}}_j] = \d 1 {l^2}\eps_{ij}^{\ \ k}
\tilde{{\cal J}}_k , \\
[\tilde{\cal H}, \tilde{{\cal K}}_i ]=0,\quad
{}[\tilde{{\cal K}}_i,\tilde{{\cal K}}_j] = 0, \quad
[ \tilde{{\cal K}}_i, \tilde{{\cal P}}_j+ \dfrac c {l^2} z^0 \tilde{{\cal K}}_j]=
\d 1 {c^2} \tilde{\cal H}\delta_{ij},\\
[\tilde{{\cal J}}_i, \tilde{{\cal J}}_j]=-\epsilon_{ij}^{\ \ k}\tilde{{\cal J}_k},
\quad [\tilde{\cal H},\tilde{{\cal J}}_i] =0, \quad
[\tilde{{\cal K}}_i,\tilde{{\cal J}}_j]=-\epsilon_{ij}^{\ \ k}\tilde{{\cal K}}_k, \\
[\tilde{{\cal P}}_i+\dfrac c {l^2}z^0\tilde{{\cal K}}_i, \tilde{{\cal J}}_j]=-\epsilon_{ij}^{\ \ k}
(\tilde{{\cal P}}_k+\dfrac c {l^2}z^0\tilde{{\cal K}}_k),
\end{cases}
\end{eqnarray}
different from
the Poincar\'e algebra.  The decomposition fits the geometrical structure on
$\mathbb{R}\times \mathbb{H}_3$.
The decomposition with the geometrical structure gives another new realization of
$\frak{iso}(1,3)$ algebra.
It is easy to see that the new realization has the space $SO(3)$ isotropy and is
invariant under the parity ($z^i\to -z^i$) and time-reversal ($z^0 \to -z^0$), i.e.
\be
\Pi:\quad  \tilde{\cal H}\to \tilde{\cal H}, \tilde{\vect{\cal P}}\to -\tilde{\vect{\cal P}},
\tilde{\vect{\cal J}}\to\tilde{\vect{\cal J}}, \tilde{\vect{\cal K}}\to
-\tilde{\vect{\cal K}}, \\
\Theta:\quad \tilde{\cal H}\to - \tilde{\cal H}, \tilde{\vect{\cal
P}}\to \tilde{\vect{\cal P}}, \tilde{\vect{\cal
J}}\to\tilde{\vect{\cal J}}, \tilde{\vect{\cal K}}\to
-\tilde{\vect{\cal K}}.%
\ee %
In addition, each $\tilde{\cal
K}_i\ (i=1,2,3)$ generates a noncompact subgroup. In other words,
the generator set $(\tilde{\cal H}, \tilde{\vect{\cal
P}},\tilde{\vect{\cal K}}, \tilde{\vect{\cal J}})$ satisfies all
three assumptions in Ref. \cite{BLL}.

 Since the algebra relation is the same as the
para-Poincar\'e algebra if $\tilde{\vect{\cal P}}$ are replaced by
$-\tilde{\vect{\cal P}}$ \cite{BLL}, the new realization of $\frak{iso}(1,3)$ algebra
is actually the para-Poincar\'e algebra.

In brief, the Beltrami translations on the 3d manifolds are
different from the algebraic (pseudo) space translations $\vect{P}'$ assigned {\it a priori}. In the new sets of generators fitting the
geometrical structure, the space-time $SO(1,3)$ isotropy
and even space $SO(3)$ isotropy are absent.  Based on the above analysis,%
\omits{The isotropic group at each
point on the manifold is $ISO(1,2)$ for the case $x\cdot x<0$ and $ISO(3)$ for
the case $x\cdot x >0$, instead of the algebraically isotropic group $SO(1,3)$.}
the space-times are the homogeneous spaces, respectively,
\be
&& M_+^{\frak{p}_2}=ISO(1,3)/ISO(1,2), \qquad x\cdot x <0, \\
&& M_-^{\frak{p}_2}=ISO(1,3)/ISO(3), \qquad \quad x\cdot x >0.
\ee
\omits{Obviously, the space-time with $x\cdot x>0$ is space isotropic in geometry.
It is a new geometry satisfying all three assumptions in Ref.\cite{BLL}.}


\section{Motions on the geometry}

Since the second Poincar\'e symmetry is found based on the
$PoR_{c,l}$ \cite{GWZ, GHWZ}, the motion for free particles should be uniform
rectilinear. In this section, we shall confine ourselves in the 4d degenerate
space-time $(M^{\frak{p}_2}_-,\vect{g}^-,\vect{h}_-,\del^-)$ and
study the motion of free particles in it, because it possesses the space
isotropy.
\omits{We find
that on the space-time $R\times H_3$, one can establish the kinematics and dynamics of
a free particle.}

\subsection{Geodesic equation}

The geodesic equation
\be
\d {d\,^2 x^\mu}{d\la^2}+\Gamma^\mu_{\
\nu\la}\d{dx^\nu}{d\la}\d{dx^\la}{d\la}=0,
\ee
gives rise to
\be
\d {d}{d\la} \left (\d 1 {x \cdot x} \d{dx^\mu}{d\la}\right )=0.
\ee
It solutions is
\be
\d 1 {x \cdot x} \d{dx^\mu}{d\la}=C^\mu.
\ee
Therefore,
\be
\d{dx^i}{dx^0}=\d {C^i}{C^0}, \qquad\Rightarrow\qquad
 x^i= a^i x^0 + lb^i, \label{gim}
\ee
where $ a^i=C^i/C^0$ and $b^i$ are two dimensionless constants.
In other words, if the motion for free particles is still determined by
the geodesic equation, the motion for free particles is a `uniform
rectilinear motion' as required if $x^0/c$ is interpreted
as the time and $x^i$ are interpreted as coordinates of space.
$ca^i$ and $lb^i$ play the roles of the uniform velocity and the initial position
respectively.
However, Eq. (\ref{gim}) reads%
\be %
z^i= b^i z^0 + la^i
\ee
in terms of the Beltrami coordinates on the $\mathbb{H}_3$ space, $z^i=l
{x^i}/{x^0}$, and $z^0=l^2/x^0$ introduced in the previous section. %
Again, it has the form of `uniform rectilinear motion' if $z^0/c$ is interpreted as the
time coordinates.  But now, $cb^i$ and $la^i$ play the roles of the uniform velocity
and the initial position, respectively.  The discrepancy raises a question: which is the
genuine velocity of the free particle moving in the space-time?

It should be noted that the geometric structure of the space-time shows that
$x^0/c$ and $z^0/c$ are not the coordinate of time and that $x^i$ are not the coordinates
of space with respect to the degenerate metric.  In the space-time, $\eta/c$ is the absolute
time and $z^i$ is the coordinates of the space $\mathbb{H}_3$.  In terms of $\eta$ and
$z^i$, Eq. (\ref{gim}) reads
\be
z^i= \d {b^i} {\cosh (r/l)} \eta + la^i,
\ee
where $r=l\tanh^{-1}(\sqrt{(z^1)^2+(z^2)^2+(z^3)^2}/l)$.  When $r\ll l$, it reduces to
$z^i= b^i \eta + la^i$.  This is a uniform rectilinear motion in the conventional sense.
Therefore, $cb^i$ is the genuine velocity of the free particle moving in a neighborhood of
the origin of the space.

\subsection{Formal Lagrangian and Euler-Lagrangian equation for a free particle}

Consider the Lagrangian
\be \label{Lagrangian}
L= \frac 1 {x \cdot
x}\sqrt{(\eta_{\mu\nu}\eta_{\rho \tau}-\eta_{\mu \rho}\eta_{\nu
\tau} )x^\rho x^\tau \dot x^\mu \dot x^\nu},
\ee
with $\dot x^\mu :={dx^\mu}/{d\la}$, where $\la$ is the affine parameter
along the trajectory of a particle.
The
Euler-Lagrangian equation reads
\be
\d d {d\la} \d {(x\cdot \dot
x) x_\ka-(x\cdot x) \dot x_\ka } {(x\cdot x)\sqrt{(x\cdot x) (\dot
x\cdot \dot x)-(x\cdot \dot x)^2}} - \d {2(x\cdot \dot
x)^2x_\ka-(x\cdot x) (\dot x\cdot \dot x)x_\ka -(x\cdot x)(x\cdot
\dot x)\dot x_\ka} {(x\cdot x)^2\sqrt{(x\cdot x) (\dot x\cdot \dot
x)-(x\cdot \dot x)^2}} =0.
\ee
After some manipulation, it reduces
to
\be
\hspace{-0.8cm}[(x\cdot x)(\dot x\cdot \dot x)-(x\cdot \dot x)^2 ]
\ddot x_\ka\ +(\dot x\cdot \ddot x)[(x\cdot \dot x) x_\ka -(x\cdot x)
\dot x_\ka] +(x\cdot \ddot x)[(x\cdot \dot x)\dot x_\ka
-(\dot x\cdot \dot x)x_\ka]=0.
\ee
This is a system of homogeneous equations for $\ddot x$.  Since its coefficient determinant
\be
\left |[(x\cdot x)(\dot x\cdot \dot x)-(x\cdot \dot x)^2 ]\dl^\la_\ka
+ \dot x^\la[(x\cdot \dot x) x_\ka -(x\cdot x)
\dot x_\ka]+x^\la[(x\cdot \dot x)\dot x_\ka
-(\dot x\cdot \dot x)x_\ka] \right |.
\ee
is not equal to 0, it has only zero solution $\ddot x_\ka= 0.$
It is equivalent to $\dot x^\ka =const.$
Thus,
\be
\d {dx^\ka}{dx^0}=const.
\ee
In other words, the generalized inertial motion can be obtained from the Lagrangian.


\section{Concluding remarks}

Bacry and L\'evy-Leblond focus their attention on the algebraic
relation in \cite{BLL}. Their theorem says that under the three
assumptions there exist only 11 kinds of kinematical algebraic
relations.  If the third assumption is relaxed, 3 kinds of
geometrical algebraic relations will be added.  In comparison, the
approach from the principle of relativity with two universal
constant, $PoR_{c,l}$, not only the algebraic relations but also
the realization of the generators are concerned. Therefore,
more possible kinematics than Bacry and L\'evy-Leblond revealed are
obtained.  The kinematics with the second Poincar\'e symmetry is one
of them. Obviously, the second Poincar\'e algebra is
isomorphic to the ordinary Poincar\'e algebra algebraically, but the
geometric realization of the two algebras are different. The second
Poincar\'e group no longer preserves the metric of Minkowski
space-time, but preserves the (non-vanishing-identically) geometry
$(M^{{\frak p}_2},\vect{g}, \vect{h},\del)$.

The geometrical analysis will, no doubts, provide a new view on
all possible kinematics. In the algebraic analysis, $H$, $H'$, and
$H^\pm$ take the role of the time translations, and $\vect{P}$,
$\vect{P}'$, and $\vect{P}^\pm$ serves as the space translations.
 The geometrical analysis, however, shows that they may have very
different meaning.  For example, in the geometry with $x\cdot
x>0$, the pseudo-space translations $\vect{P}'$ (relating to
$\tilde{\vect{\cal K}}$) actually generate the new kind of the
boost transformations on $\mathbb{R}\times \mathbb{H}_3$, while the Beltrami space
translations on the $\mathbb{H}_3$ space are generated by $\tilde{\vect{\cal
P}}$ which is proportional to the Lorentz boost $\vect{K}$.
This can be seen in another way.  In this case, we have $\mathfrak{p}_2$-invariant degenerate
metric $\vect{g}^-$ and absolute `time' direction $\partial_{\eta}$.
Because $\partial_{\eta}$ is unique and $\vect{g}^-$ is
independent of $\eta$, the manifold $M^{{\frak p}_2}$ has a line bundle structure
$\pi: M^{{\frak p}_2}\to\Si=\mathbb{H}_3, (\eta,z^i)\mapsto(z^i)$, where $\partial_{\eta}$
is the tangent direction of the fiber. The $\mathbb{R}^4$ ideal of $\frak{iso}(1,3)$
algebra are all along the fiber direction and $(\mathbb{H}_3, \vect{g}^-)$ can be
seen as an
``absolute space". Combine the ${\frak p}_2$ action on $M^{{\frak p}_2}$ and
$\pi:M^{{\frak p}_2}\to\Sigma$, we can define the ${\frak p}_2$ action on $\Sigma$ as
\[
g(\vect{z})=\pi\circ g\circ\pi^{-1}(\vect{z}),\ \forall \vect{z}\in\Sigma\ {\rm and}\ \forall
g\in P_2.
\]
Under this definition, the actions of the $\mathbb{R}^4$ ideal are trivial
on $\Sigma$, i.e. they are no longer `space translations'.  The ${\frak p}_2$
action defined above is equivalent to the ${\frak L}_p$ action on $\Sigma$.  And the three
boosts $\{K_i\}$ combined with $(1/x^0)P_i$, respectively, take the place of `space translations', like the
original space translation, spanning a representation
space of the $\mathfrak{so}(3)$ sub-algebra on $\Sigma$.

The difference between the two Poincar\'e algebras should be further remarked on.
In the above ${\frak p}_2$ algebra, the $\frak{so}(3)$ on $\Sigma$ is unique.
In contrast, in the ordinary Poincar\'e algebra, the
choice of the $\mathfrak{so}(3)$ in ${\frak L}_p$ is not canonical.  The division
of the ideal of $\frak p$ into $\mathbb{R}\oplus \mathbb{R}^3$ based on the irreducible
representation of the $\mathfrak{so}(3)$ depends on the choice. The different choice
of the $\mathfrak{so}(3)$
corresponds to different sets of inertial observers.

Like the Galilei and Carroll space-times, the space and time
of the new geometry $\{M^{\frak{p}_2},\vect{g}, \vect{h},\del\}$ are split.
For the $x\cdot x <0$ case, 1-d space is split out.  There is a special
direction in space.  The kinematics on the 3d space-time is still relativistic,
but is dramatically different from the kinematics in the special direction.
It should be noted that $z^3$ is not the intrinsic coordinates for the split-out space.
In terms of the intrinsic coordinates $z^\al, \rho$, Eq.(\ref{P3'+}) and Eq.(\ref{H'+})
become
\be
{P}'_3 = -\d 1 {\sqrt{1-l^{-2}\eta_{\al\beta}z^\al z^\beta}}\r_\rho =-{\cal P}_3,
\ee
\be
\d l {c^2}  H'=\d 1 c \d {z_0}
{\sqrt{1-l^{-2}\eta_{\al\beta}z^\al z^\beta}}\r_{\rho}={{\cal K}}_3,
\ee
When $|z^i|\ll l$, they tend to the ordinary translation ${\cal P}_3 \approx \r_\rho$
and Galilei boost ${{\cal K}}_3 \approx c^{-1}  {z_0} \r_{\rho}$, \omits{${{\cal J}}_1
\approx {z_1}\r_{\rho}$, and ${{\cal J}}_2 \approx
{z_2}\r_{\rho}$,} respectively.
For the $x\cdot x>0$, the time is split out, which fixes a special
time direction and an absolute space.  In terms of the intrinsic coordinates
$\eta, z^i$, Eq.(\ref{H'-}) and Eq.(\ref{P'-}) become, respectively,
\be
H'= \d c {\sqrt{1-l^{-2}\dl_{ij}z^i z^j}}\r_\eta =\tilde{\cal H}
\ee
and
\be
\d l c {P}'_i=\d 1 c\d {z_i}{\sqrt{1-l^{-2}\dl_{jk}z^j z^k}}\r_\eta=\tilde{ K}_i.
\ee
When $|z^j|\ll l$, they reduce to the ordinary time translation $\tilde{\cal H}\approx
c^{-1}\r_\eta$ and the Carroll boosts $\tilde{{\cal K}}_i\approx  - c^{-1}z^i\r_\eta$,
respectively.  The latter situation is very similar to
the Carroll algebra and Carroll space-time, in which there is a special time
direction and an absolute space.  The difference between the Carroll space-time
and the new space-time is that the absolute space in Carroll space-time is flat while
the absolute space in the new space-time is Lobachevskian.
In this sense, the new kinematics is non-relativistic.

If the space isotropy is required on the both algebraic
and geometrical levels, only the space-time with $x\cdot x>0$ remains.
On the new space-time, the motions of free particles can be well defined.
The mechanics, field theories and even gravity on the space-time should be further
investigated
in order to clarify the application of the new space-time.  In the higher dimensional
theories, there may be the second Poincar\'e group as its subgroup of symmetry.  Hence,
the geometric structure may appear in a higher dimension.

The reason that only the geometries for $x\cdot x<0$ and $x\cdot x>0$ cases are
presented is that $x\cdot x=0$ defines a three dimensional hypersurface,  while
the possible kinematics we are interested in is defined on a 4-d manifold.

\begin{acknowledgments}\vskip -4mm
We are very grateful to Prof. H.-Y. Guo for his valuable suggestions
and comments.  We would also like to thank
Z.-N. Hu, W.-T. Ni and Dr. H.-T. Wu for their
helpful discussion.
 This work is supported by NSFC under Grant Nos. 10775140,
10705048, 10731080, 10975141, the President Fund of
GUCAS, and the Fundamental Research Funds for the Central Universities
under Grant No. 105116.
\end{acknowledgments}

\end{CJK*}
\end{document}